\documentclass[twocolumn,superscriptaddress,prx]{revtex4}
\usepackage{graphicx}

\usepackage{multirow}
\usepackage{amssymb}
\usepackage{graphicx}
\usepackage{amsmath}
\usepackage{graphicx}
\usepackage{dcolumn}
\usepackage{bm}
\usepackage{color}
\usepackage{float}

\usepackage{setspace}

\newcommand{\be}{\begin{equation}}
\newcommand{\ee}{\end{equation}}
\newcommand{\bea}{\begin{eqnarray}}
\newcommand{\eea}{\end{eqnarray}}

\newcommand{\degree}{\ensuremath{^\circ}}

\begin{document}

\title{Strain engineering in semiconducting two-dimensional crystals}

\author{Rafael Rold\'an}
\email{rroldan@icmm.csic.es}
\affiliation{Instituto de Ciencia de Materiales de Madrid, CSIC, Madrid, Spain}
\affiliation{Instituto Madrile\~no de Estudios Avanzados en Nanociencia (IMDEA-Nanociencia), 28049, Madrid, Spain}
\author{Andres Castellanos-Gomez}
\email{andres.castellanos@imdea.org }
\affiliation{Instituto Madrile\~no de Estudios Avanzados en Nanociencia (IMDEA-Nanociencia), 28049, Madrid, Spain}
\author{Emmanuele Cappelluti}
\affiliation{Istituto de Sistemi Complessi, U.O.S. Sapienza, CNR, v. dei Taurini 19, 00185 Roma, Italy}
\author{Francisco Guinea}
\affiliation{Instituto Madrile\~no de Estudios Avanzados en Nanociencia (IMDEA-Nanociencia), 28049, Madrid, Spain}
\affiliation{School of Physics and Astronomy, University of Manchester, Oxford Road, Manchester M13 9PL, UK}


\begin{abstract}

One of the fascinating properties of the new families of two-dimensional crystals is their high stretchability and the possibility to use external strain to manipulate, in a controlled manner, their optical and electronic properties. Strain engineering, understood as the field that study how the physical properties of materials can be tuned by controlling the elastic strain fields applied to it, has a perfect platform for its implementation in the atomically thin semiconducting materials. The object of this review is to give an overview of the recent progress to control the optical and electronics properties of 2D crystals, by means of strain engineering. We will concentrate on semiconducting layered materials, with especial emphasis in transition metal dichalcogenides (MoS$_2$, WS$_2$, MoSe$_2$ and WSe$_2$). The effect of strain in other atomically thin materials like black phosphorus, silicene, etc., is also considered. The benefits of strain engineering in 2D crystals for applications in nanoelectronics and optoelectronics will be revised, and the open problems in the field will be discussed.

\end{abstract}

\maketitle

\tableofcontents{}



\section{Introduction}

The isolation of graphene by mechanical exfoliation in 2004 \cite{A1} opened the door to study a broad family of two-dimensional materials \cite{A2,A3,A4,A5}, almost unexplored at that time. These materials are characterized by strong covalent in-plane bonds and weak interlayer van der Waals interactions which give them a layered structure. This rather weak interlayer interaction can be exploited to extract atomically thin layers by mechanical \cite{A2} or chemical exfoliation methods \cite{A6}. In the last five years a wide variety of materials with very different electronic properties (ranging from wide-bandgap insulators to superconductors) have been explored \cite{A7,A8,A9,A10,A11,A12,A13}. These materials are expected to complement graphene in applications for which graphene does not possess the optimal properties. Two-dimensional semiconductors with an intrinsic large bandgap are a good example of materials that could complement graphene \cite{A3,A14}. While the outstanding carrier mobility of graphene makes it very attractive for certain electronic applications (e.g. high-frequency electronics), the lack of a bandgap dramatically hampers its applicability in digital electronics. Graphene field-effect transistors suffer from large off-state currents due to the low resistance values at the charge neutrality point. The large off-state current also translates to a large dark current in graphene-based photodetectors which limits their performance.
The handicaps of gapless graphene-based electronic devices have motivated experimental efforts towards opening a bandgap in graphene in order to combine its excellent carrier mobility with a sizable bandgap. Several methods have been employed: lateral confinement \cite{A15,A16,A17}, application of a perpendicular electric field in bilayer graphene \cite{A18,A19}, hydrogenation \cite{A20,A21,A22,A23}, or controlled nanoperforation \cite{Pedersen08,Kim10,Yuan13} among others. However, in all cases the magnitude of the bandgap is not enough to ensure high-performance room-temperature and/or the mobility of the treated graphene severely decreases.

The difficulty of achieving high-mobility gapped graphene has triggered the study of semiconducting analogues to graphene in materials with an intrinsic bandgap \cite{A3,A24,A25,A26,A27,A28}. For instance single-layer MoS$_2$, the most studied 2D semiconductor so far, presents a direct bandgap (1.8 eV) and a large in-plane mobility (up to 200 cm$^2$V$^{-1}$s$^{-1}$) \cite{A14,A29,A30,A31,A33,A34,A35}. The intrinsic bandgap of MoS$_2$ is essential for many applications, including transistors for digital electronics or certain optoelectronic applications. For instance, field-effect transistors based on single-layer MoS$_2$ showed room-temperature current on/off ratios of $10^7-10^8$ and ultralow standby power dissipation \cite{A14}. Logic circuits, amplifiers and photodetectors based on monolayer MoS$_2$ have also been recently demonstrated \cite{A33,A36,A37}.

Apart from their outstanding figures-of-merit and electrical performances, atomically thin semiconductors have also shown very interesting mechanical properties \cite{CS15}, unmatched by conventional 3D semiconductors \cite{A38,A39}. Nanoindentation experiments with an atomic force microscope tip on freely suspended single-layer MoS$_2$ have shown mechanical properties approaching those predicted by Griffith for ideal brittle materials in which the fracture point is dominated by the intrinsic strength of their atomic bonds and not by the presence of defects \cite{A40}. While the ideal breaking stress value is expected to be one ninth of the YoungÕs modulus, for single layer MoS$_2$ it reaches $\sim 1/8$ \cite{A38,A41}. This almost ideal behavior is attributed to a low density of defects on the fabricated devices, probably due to the lack of dangling bonds or other surface defects and their high crystallinity. This is in clear contrast with conventional 3D semiconductors. Indeed, while silicon typically breaks at strain levels of $\sim1.5$\%, MoS$_2$ does not break until $>10\%$ strain levels and it can be folded and wrinkled almost at will \cite{A42,A43}. On the other hand, the elastic properties of 2D crystals at long wavelengths are dominated by anharmonic out-of-plane vibrations, and a controlled creation of vacancies \cite{LG15} or strain \cite{RK10,LopezPolin15} can be used to enhance the strength of these materials. 
Finally, the ultimate crystal instability (onset of negative phonon frequencies) of several families of two-dimensional materials
was investigated in Ref. \cite{isaacs}.

This outstanding stretchability of 2D crystals promises to revolutionize the field of strain engineering due to the unprecedented tunability levels predicted for these materials that could lead to "straintronic" devices Ð devices with electronic properties that are engineered through the introduction of mechanical deformations.
Single-layer MoS$_2$, for instance,
is expected to undergo a direct-to-indirect bandgap transition
at $\sim 2$\% of tensile uniaxial strain, and a semiconducting-to-metal transition at $10-15\%$ of tensile biaxial strain \cite{A44,A45,A46}.
Such huge sensitivity of the bandgap (from 1.8 eV  to 0 eV)
can be realistically exploited in MoS$_2$ and other transition metal dichalcogenides, since these materials can withstand such levels of strain without rupture \cite{A42,A43}. This behavior has to be compared with the poor window of tunability of only 0.25 eV achieved for strained silicon at 1.5\% biaxial strain, due to the breaking of the bonds beyond that value of strain \cite{A47}.
Another game-changing feature of atomically thin semiconductors is that one can deform them only at a specific location making it possible to generate localized strain profiles. One possible advantage of local strain engineering has been recently proposed by Feng {\it et al.} in Ref. \cite{A45} when a MoS$_2$ membrane is subjected to a point deformation at its centre, a continuous bending of the energy levels of electrons, holes and excitons is produced. Upon illumination electron-hole pairs are generated and the semiclassical potential for the quasiparticles exerts forces on the electron and holes in opposite directions. However, the large exciton binding energy will force the electron-hole pairs to hold together and migrate towards the centre of the membrane. Therefore, localized strains can be used to concentrate excitons in a small region of the semiconductor crystal with interest for fundamental physics (e.g. exciton condensates) and applications (e.g. solar energy funnel).

The present review covers the different levels in the study of strain engineering in 2D crystals, starting from the fabrication in Sec. \ref{Sec:Fabrication}. The different techniques to induce strain are discussed in Sec. \ref{Sec:Techniques} and their effects on the optical and electronic properties are summarized in Sec. \ref{Sec:BandGap}, where we list the main experimental results. From a theoretical point of view, the electronic structure of semiconducting 2D materials is revised in Sec. \ref{Sec:NoStrain}. In Sec. \ref{Sec:Homogeneous} we go over the effect of homogeneous strain on these materials, focusing on transition metal dichalcogenides, black phosphorus and silicene, and the effect of non-uniform strain is discussed in Sec. \ref{Sec:Inhomogeneous}. In Sec. \ref{Sec:Conclusions} we give our conclusions and final remarks.

\section{Two dimensional crystals}\label{Sec:Experiment}

\subsection{Fabrication}\label{Sec:Fabrication}

In this section the main fabrication methods to isolate or synthesize 2D materials will be briefly described. For a more detailed discussion about fabrication methods we address the reader to recently published review articles that are focused on this specific topic \cite{A48,A49}. Different fabrication techniques typically yield different size, thickness and even quality of the fabricated 2D materials and therefore the selection of the isolation method strongly depends on the desired application. 

\begin{figure}[t]
\includegraphics[scale=0.9,clip=]{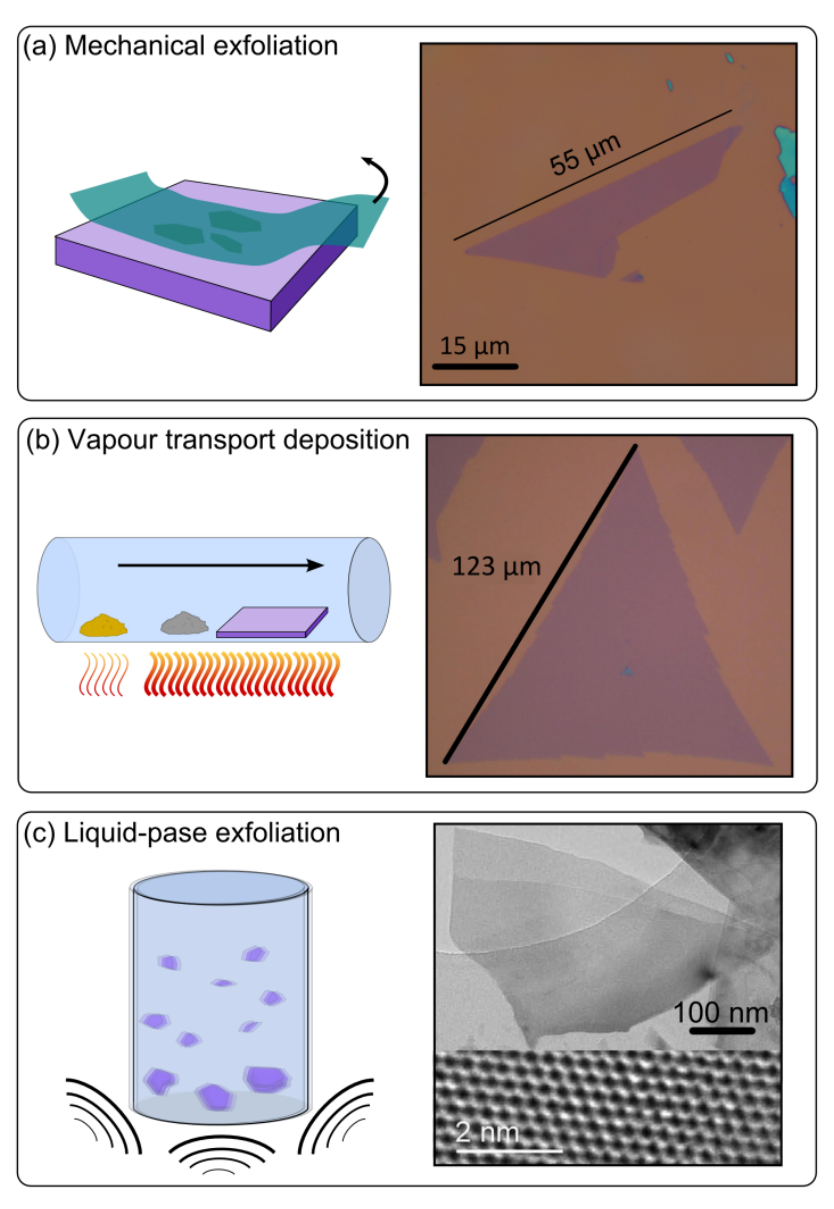}
\caption{
{\bf Isolation of single-layer MoS$_2$ by different techniques.}  (a) Optical image of a MoS$_2$ flake deposited on a SiO$_2$/Si substrate by mechanical exfoliation. (b) Optical image of a single-layer MoS$_2$ crystal grown on a SiO$_2$/Si by chemical vapour deposition method (adapted from Ref. \cite{A57} with permission of Nature Publishing Group). (c) Transmission electron microscopy image of a MoS$_2$ flake isolated by ultrasonication of MoS$_2$ powder in an organic solvent. The inset shows a high-resolution scanning transmission electron microscopy image of the crystal lattice of the single layer region. Adapted from Ref. \cite{A65} with permission of Science.
}
\label{Fig:Fig1}
\end{figure}

\begin{figure*}[t]
\includegraphics[scale=1.0,clip=]{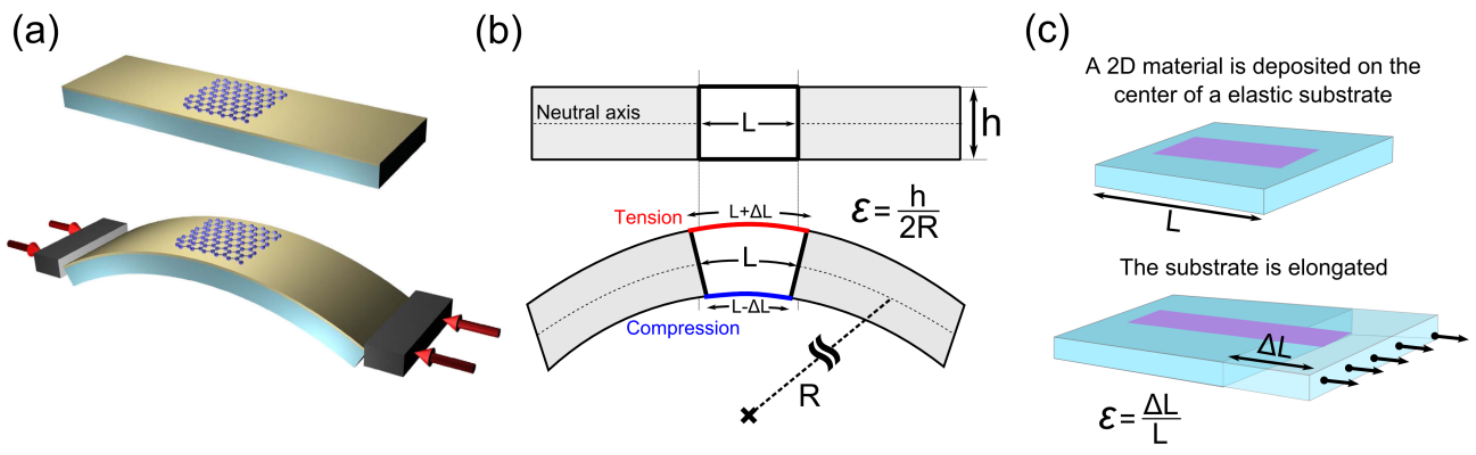}
\caption{
{\bf Uniaxial straining of 2D materials with bending and tensile test geometry.}   (a) Cartoon of a flexible substrate with a 2D material deposited onto its surface. By using two-point bending test geometry, the substrate can be bent to apply a controlled uniaxial tensile strain to the 2D material. (b) Simplified scheme of the cross section of the flexible substrate before and after bending it. (c) Scheme of the approach to apply uniaxial tensile strain by elongating an elastic substrate. Panel (a) adapted from \cite{A72} with permission of the American Physical Society.
}
\label{Fig:Fig2}
\end{figure*}

\subsubsection{Mechanical exfoliation}

The weak van der Waals interaction between the layers makes it possible to cleave thin crystalline flakes by peeling off the surface of a bulk layered material that is adhered to a piece of sticky tape. This method to isolate thin sheets from a bulk layered material is referred to in the literature as ÔScotch tape methodÕ, mechanical exfoliation or micromechanical cleavage and it has proven to be a simple yet powerful technique to obtain high-quality two-dimensional sheets. The process is based on repeatedly cleaving a bulk layered material with a sticky tape to generate a large quantity of thin crystallites \cite{A1}. These crystallites can be transferred to an arbitrary flat substrate by gently pressing the tape against the surface of the substrate and peeling it off slowly. More details on this technique can be found in the pioneering work of Novoselov, Geim {\it et al.} \cite{A2}.

The main limitation of this technique is the fact that during the deposition step flakes with various thicknesses are transferred all over the surface and only a small fraction of them are few-layers thick. Nonetheless, this issue can be overcame to a large extent because optical microscopy can be used to identify thin 2D flakes in a fast and reliable way \cite{Blake07}. Fig. \ref{Fig:Fig1}(a) shows an example of optical microscopy image of a single-layer MoS$_2$ flake. Its thickness can be distinguished at glance by its optical contrast, due to a combination of interference color and optical absorption. Several quantitative studies of the optical contrast of many 2D materials, providing a guide to optically identify 2D materials, have been recently reported: MoS$_2$ \cite{A9,A50,A51,A52}, NbSe$_2$,\cite{A9,A52}, WSe$_2$ \cite{A51,A52}, TaS$_2$ \cite{A51}, TaSe$_2$ \cite{A53}, mica \cite{A11,A54}.

\subsubsection{Chemical vapour deposition or epitaxial growth}

2D materials can be also prepared by bottom-up approaches where layers are synthesized by assembling their elemental constituents. Transition metal dichalcogenides (TMDs) have been synthesized by vapour transport method \cite{A55,A56}, chemical vapour deposition \cite{A57,A58,A59,A60} or van der Waals epitaxy \cite{A61,A62,A63}. High-quality molybdenum and tungsten dichalcogenides can be directly grown on insulating substrates with large single-crystal domains of 10 to 10000 $\mu$m$^2$ \cite{A56,A57,A60}. Fig. \ref{Fig:Fig1}(b) shows an optical microscopy image of a typical single-crystalline MoS$_2$ monolayer grown by chemical vapour deposition \cite{A57}.

\subsubsection{Liquid-phase exfoliation}

Alternatively to the approaches described below, atomically thin 2D crystals can be also isolated by exfoliating bulk layered materials immersed in a liquid medium using two main approaches: direct sonication in a solvent or intercalation/expansion/sonication. We address the reader to Refs. \cite{A64,A65,A66,A67,A68} and to Refs. \cite{A69,A70} for details about these two liquid phase exfoliations approaches, respectively.

Due to the combination of high yield and low cost, this technique is a prospective fabrication approach to fabricate large quantities of atomically thin crystals but it yields small size flakes ($0.2-1 \mu{\rm m}^2$) and it lacks of sufficient control to achieve monodisperse solutions. Note that liquid phase exfoliated flakes have not been used in the strain engineering studies reported so far, probably because of their reduced lateral sizes. Fig. \ref{Fig:Fig1}(c) shows a transmission electron microscopy image of a MoS$_2$ flake fabricated by liquid phase exfoliation of MoS$_2$ powder \cite{A65}.

\subsection{Techniques to induce strain}\label{Sec:Techniques}

In this section the methods employed to strain engineer 2D materials will be described.
Leaving aside techniques which have been employed for specific purposes in graphenes
and not yet applied for the families of 2D compounds we focus here (transition metal dichalcogenides,
black phosphorus, silicenes, \ldots), we will address in the following four techniques
that have already proven to be feasible and currently employed for these materials:
bending of flexible substrates, elongating an elastic substrate, piezoelectric stretching, substrate thermal expansion and controlled wrinkling. These methods will be described in separated subsections. A summary of this section will compare the results obtained by the different straining approaches.  

\subsubsection{Bending of a flexible substrate}

Controllable uniaxial tensile strain can easily be applied to 2D materials by using a modified version of a bending test. The 2D material is deposited onto a flexible substrate (directly by mechanical exfoliation of a bulk layered material or by transferring a vapour-transport growth film). By bending the flexible substrate as depicted in Fig. \ref{Fig:Fig2}(a) the topmost surface is stretched, transferring a uniaxial tensile strain to the 2D material that lays on the surface. The bending of the flexible substrate is typically carried out with a micrometer stage in a two-points, three-points or four-points bending geometry. 

The strain achieved by bending the flexible substrate can be obtained from a continuum mechanics model for elastic beams assuming that the radius of curvature ($R$) due to bending the substrate is larger than the substrate thickness ($h$)  \cite{A71} (see sketch in Fig. \ref{Fig:Fig2}(b)). Also, the dominant deformation of the beam should be in the longitudinal direction. This means that shear stresses and stresses normal to the neutral axis are negligible. Based on these assumptions the strain can be obtained as \cite{A71}:
\begin{equation}
\varepsilon=h/2R
\end{equation}

The van der Waals interaction between the 2D material and the surface of the flexible substrate is typically strong enough to clamp the nanosheet and prevent slippage for strain levels up to 1\%. For higher strain levels, the sheets tend to slip and the tension is released hampering the acquisition of reproducible datasets. In order to reduce the slippage issue, one can evaporate metal strips onto the nanosheets to act as clamping points \cite{A73}.

This technique to apply strain has been successfully employed to study the changes in the Raman vibrational modes of graphene subjected to tensile strain \cite{A72,A74,A75,A76}. More recently, this straining technique has been adopted to study the role of uniaxial deformation in the optical properties of atomically thin MoS$_2$ which will be described in the next section \cite{A77,A78,A79}. The maximum reported uniaxial strain achieved by this technique ranges from 0.5 \% to 2.5\%, depending on the exact experimental conditions.

\begin{figure}[t]
\includegraphics[scale=0.90,clip=]{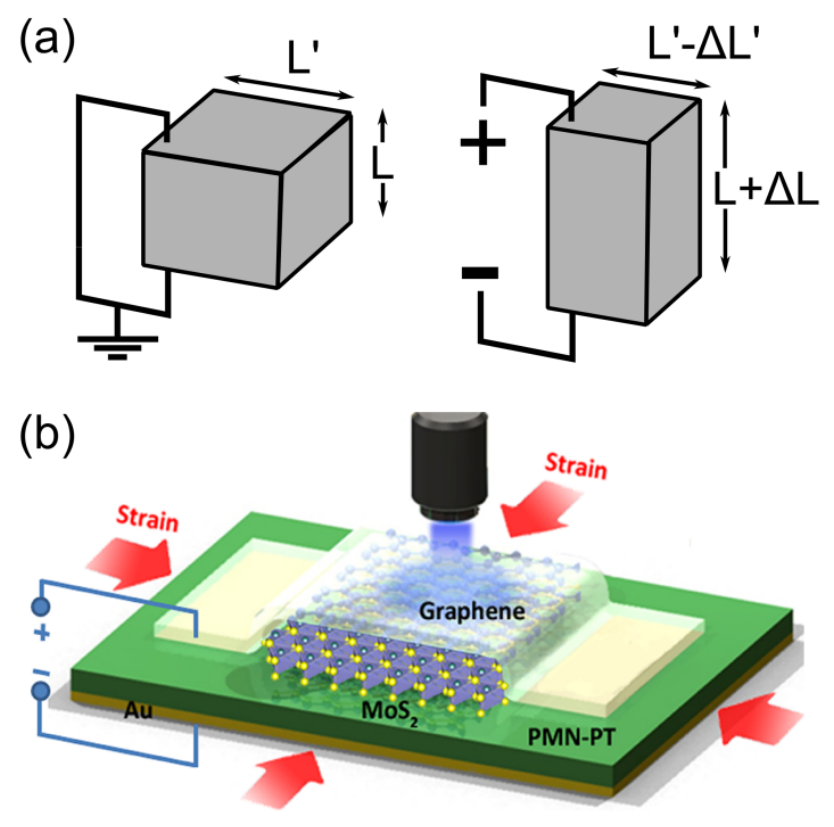}
\caption{
{\bf Uniform biaxial straining by using piezoelectric substrates.}    (a) Schematic diagram of the mechanical deformation of a piezoelectric material when an electric field is applied. (b) Cartoon of the experimental setup employed in Ref. \cite{A81} to apply uniform biaxial strain to trilayer MoS$_2$ (adapted with permission of the American Chemical Society).
}
\label{Fig:Fig3}
\end{figure}

\subsubsection{Elongating the substrate}\label{Sec:Elongatation}

Alternatively to the previous technique homogeneous uniaxial tensile strain can be also applied by depositing the 2D material on an elastic substrate and elongating it with a straining stage, like the ones used for tensile tests in material science (see Fig. \ref{Fig:Fig2}(c)). This method has been recently used to apply uniaxial tensile strain (up to 4.0\%) to monolayer WS$_2$ \cite{A80}.

\subsubsection{Piezoelectric stretching}

Piezoelectric materials are materials that respond to external electric fields with a mechanical deformation. Fig. \ref{Fig:Fig3}(a) shows a schematic diagram of a piezoelectric material that is elongated in the vertical direction and shrink in the horizontal one when an electric field is applied (by applying a bias voltage between two faces of the piezoelectric material). Therefore, the use of piezoelectric substrates is very appealing to apply controllable strains to 2D materials as one can easily regulate the deformation of the substrate by simply varying the applied voltage to the piezoelectric material.  

Hui {\it et al.} have transferred trilayer MoS$_2$ onto a [Pb(Mg$_{1/3}$Nb$_{2/3}$)O$_3$]$_{0.7}$-[PbTiO$_3$]$_{0.3}$ piezoelectric substrate \cite{A81}. By applying a bias voltage to the substrate (as illustrated in the sketch in Fig. \ref{Fig:Fig3}(b)) the surface where the flake is deposited shrink, subjecting the trilayer MoS$_2$ to a uniform compressive biaxial strain. With this technique only moderate compressive biaxial strain levels have been achieved (a maximum 0.2 \% have been reported so far). Note, however, that biaxial strain has a stronger effect on the band structure of transition metal dichalcogenide materials than uniaxial strain (see Sec. \ref{Sec:Biaxial}).

\subsubsection{Exploiting the thermal expansion mismatch}

\begin{figure}[t]
\includegraphics[scale=0.90,clip=]{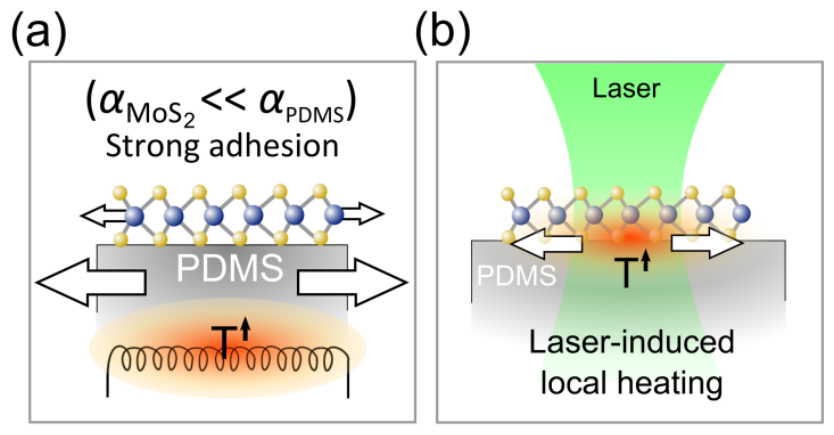}
\caption{
{\bf Uniform biaxial straining by exploiting the thermal expansion of the substrate.}     (a) Schematic diagram of the process to strain atomically thin materials by using substrates with strong adhesion and very large thermal expansion coefficient. (b) Cartoon of the straining process using a focalized laser as a heating source. Adapted from Ref. \cite{A83} with permission of the Institute Of Physics.
}
\label{Fig:Fig4}
\end{figure}

Thermal expansion mismatch is typically seen as a ÔdrawbackÕ in microfabrication of thin film devices process as it leads to undesired strain during the different thermal processes needed for the fabrication of micro-devices. One can, however, exploit the thermal expansion mismatch to engineer strain on a semiconducting material in order to modify its optoelectronic properties in a controllable way.
The sketch in Fig. \ref{Fig:Fig4}(a) shows an example of how a large mismatch between the thermal expansion of a 2D semiconducting material and of the substrate where it is laying can be exploited to induce uniform biaxial tensile strain. While 2D materials typically have negative thermal expansion coefficients, the substrate can be intentionally selected to have a very large positive thermal expansion coefficient. Therefore, by warming up the substrate the 2D material will experience a tensile biaxial strain.

This effect has been observed by Plechinger {\it et al.} in single-layer MoS$_2$ sandwiched between two dielectric layers \cite{A82}. By changing the sample temperature a shift in the photoluminescence spectra evidenced a change in the electronic band structure that can be explained by the tension induced during the thermal cycling by the mismatch in thermal expansion coefficients. In these experiments, however, the magnitude of the induced strain was modest because of the relatively small thermal expansion coefficient of the dielectric layers employed. More recently, it has been shown how this effect can be magnified by selecting a substrate with a large thermal expansion coefficient. For example, by using poly-dimethil siloxane instead of conventional dielectric substrates it has been recently achieved 0.23 \% of biaxial tensile strain by increasing the temperature of the sample by 150 \degree C \cite{A83}. In that reference, they also demonstrated that this technique can be modified to attain non-homogeneous strain profiles. In fact Plechinger {\it et al.} proposed to employ a focalized laser beams to locally increase the temperature of the substrate at certain positions which would lead to more complex straining patterns \cite{A83}.

\subsubsection{Controlled wrinkling}

\begin{figure*}[t]
\includegraphics[scale=0.90,clip=]{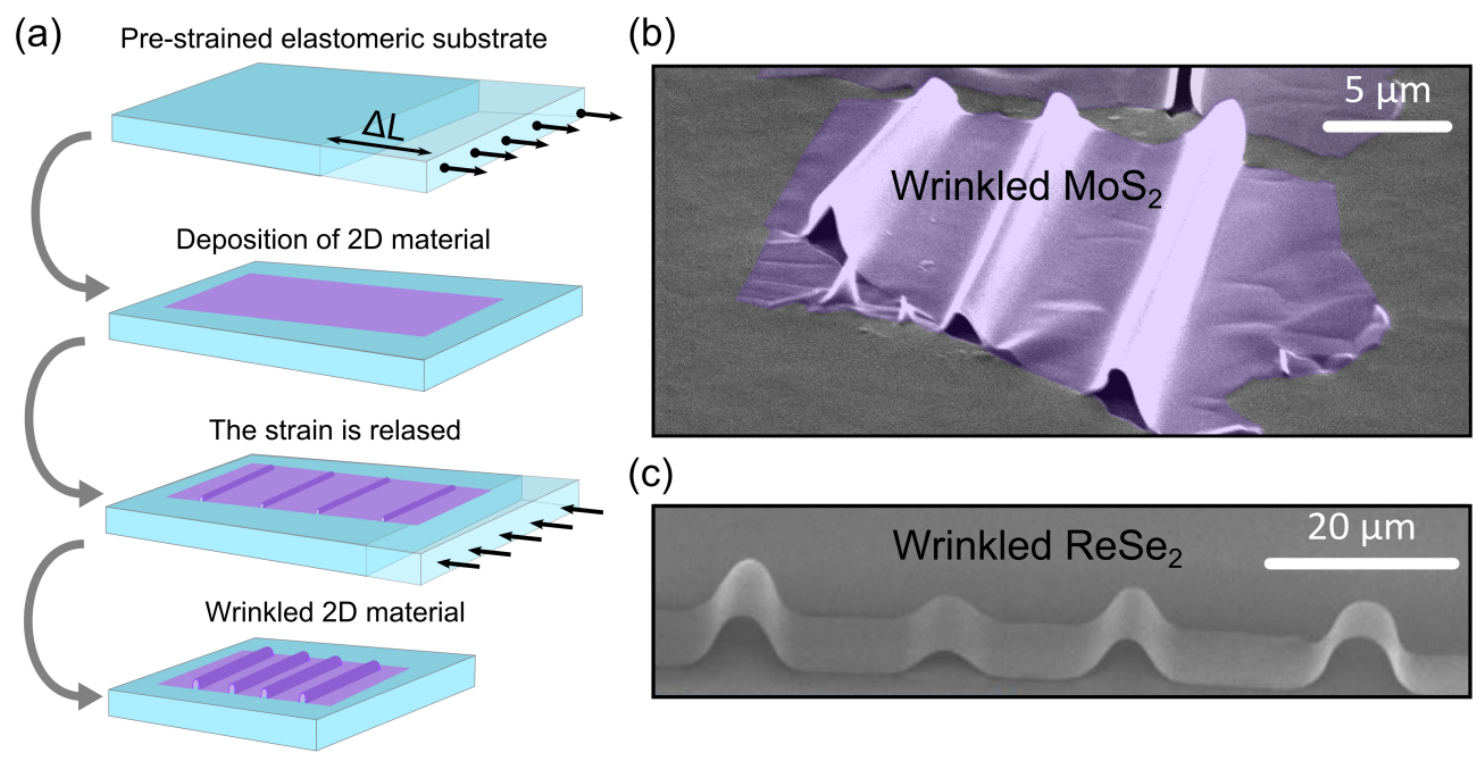}
\caption{
{\bf Non-homogeneous straining by wrinkling 2D materials.}      (a) Schematic diagram of the fabrication process of wrinkled nanolayers. An elastomeric substrate is stretched prior depositing the 2D material by mechanical exfoliation. The strain is then released producing buckling-induced delamination of the flakes. (b) and (c) High angle scanning electron microscopy images of wrinkled MoS$_2$ and ReSe$_2$, respectively. Panel (a) adapted from Ref. \cite{A84} with permission of the American Chemical Society. Panels (b) and (c) are adapted from Ref. \cite{A84} and Ref. \cite{A85} respectively with permission of the American Chemical Society.
}
\label{Fig:Fig5}
\end{figure*}

Transferring 2D materials onto elastomeric substrates allows one to  generate non-homogeneous deformations in a relatively simple way by exploiting the phenomena known as buckling-induced rippling or wrinkling. Fig. \ref{Fig:Fig5}(a) shows a schematic diagram of the buckling-induced rippling/wrinkling process. A thin film is deposited onto an elastomeric substrate that has been uniaxially deformed prior to the deposition step. Once the thin film is deposited the flexible substrate strain is released which lead to a uniaxial compression of the thin film. If the initial strain is below a critical value, the thin film will deform upon the compressive strain in a sinusoidal way without delaminating from the substrate (rippling). For higher initial strain levels, the thin film delaminates from the substrate forming wrinkles that accumulate the deformation, separated by flat unstrained regions. This process has been previously used to generate periodic ripples in graphene \cite{Wang11,Zang13} and it has been recently used to generate non-homogeneous strain profiles in atomically thin MoS$_2$ and ReSe$_2$ \cite{A84,A85}. Fig. \ref{Fig:Fig5}(b) and (c) show high angle scanning electron microscopy images of wrinkles generated in thin MoS$_2$ and ReSe$_2$ layers using the buckling-induced delamination process. This method provides a powerful, yet simple, way to achieve strongly localized strains up to a 2.5\% \cite{A84}. As it will be discussed in the coming sections, localized strains can be exploited to collect photogenerated charge carriers by generating an exciton funnel.

The maximum local strain $\varepsilon$ is induced on the topmost part of the winkles and it can be estimated by the following formula $\varepsilon\sim\pi^2h\delta/(1-{\tilde\nu}^2)\lambda^2$ \cite{A84,A86}, where $\tilde\nu$ is the PoissonÕs ratio, $h$  is the thickness of the flake, and $\delta$ and $\lambda$ are the height and width of the wrinkle extracted from atomic force microscopy. Alternatively to the strain estimation from the geometry of the wrinkles, the maximum strain level can also be obtained from the shift of the E$^1_{2g}$ Raman peak. In fact, Rice {\it et al.} studied the effect of uniaxial tensile strain on single-layer MoS$_2$, finding that the E$^1_{2g}$ Raman peak is shifted towards lower Raman shifts with uniaxial tensile strain (it shifts by $-1.7 {\rm cm}^{-1}$ per \%) while the A$_{1g}$ peak is almost strain independent \cite{A87}.

\begin{table*}
\begin{tabular}{|p{4.2cm}|c|c|c|c|c|}
\hline
\hline
{\bf Straining technique}  &\multicolumn{2}{c}{{\bf Type of strain}}& {\bf Max. strain} & {\bf Material} & {\bf Ref.}\\
\hline
\multirow{4}{4.2cm}{Bending of a flexible substrate}  & \multirow{4}{*}{Uniaxial} &   \multirow{4}{*}{Homogeneous} & 2.4\%  & MoS$_2$ & \cite{A77} \\
  & & & 0.5\% & MoS$_2$ & \cite{A78}  \\
  & & & 0.8\%  & MoS$_2$ & \cite{A79} \\
  & & &  2.1\% & WSe$_2$ & \cite{A88} \\
\hline
Elongating the substrate & Uniaxial & Homogeneous & 4.0\% & WS$_2$ & \cite{A80} \\
\hline
Piezoelectric stretching & Biaxial & Homogeneous & 0.2\% & MoS$_2$ & \cite{A81} \\
\hline
Exploiting the thermal  expansion mismatch & Biaxial & Homogeneous & 0.23\% & MoS$_2$ & \cite{A83}\\
\hline
\multirow{2}{4.2cm}{Controlled wrinkling} & \multirow{2}{*}{Uniaxial} & \multirow{2}{*}{Inhomogeneous} & 2.5\% & MoS$_2$ & \cite{A84} \\
& & & 1.6\%  & ReSe$_2$ & \cite{A85}\\
\hline
\end{tabular}
\caption{Comparison between the different techniques to apply strain. The corresponding bibliographic reference is given in square brackets in the last column.}
\label{Tab:Strains}
\end{table*}

\subsubsection{Comparison between different techniques to induce strain}

In this section, we summarize the different straining techniques described in the previous sections in order to facilitate a comparison by the reader. This summary  allows one to find the most adequate straining technique for each application. Table \ref{Tab:Strains} displays useful information for each straining method such as the type of strain achieved (uniaxial/biaxial and homogeneous/inhomogeneous), the maximum strain level reported for each technique in the literature and the 2D materials that have been strain engineered using these methods.

\subsection{Bandgap engineering}\label{Sec:BandGap}

In this section the recent experimental progress on modifying the band structure of atomically thin semiconducting materials, beyond graphene, by strain engineering will be reviewed. 
The first experimental works that demonstrate the control over the optoelectronic properties of atomically thin semiconducting materials are relatively young and are almost limited to MoS$_2$. In 2012, Plechinger {\it et al.} reported a shift of the photoluminescence yield of single-layer MoS$_2$ sandwiched between two dielectric layers when the temperature of the stack was cycled between 4 K and 240 K \cite{A82}. Similar results were found by Yan {\it et al.} \cite{A89} for single-layer MoS$_2$ samples also sandwiched between two dielectric layers and heated with an increasingly high power laser. These results were understood in the basis that the thermal expansion coefficient mismatch between the MoS$_2$ and the surrounding dielectric layer generated a biaxial strain when the temperature of the stack was varied. In these works, however, the magnitude of the applied strain was modest and the maximum attained strain level was not quantified. Soon after these works, five different groups simultaneously reported their experimental result demonstrating the control over the band gap of atomically thin MoS$_2$ by using mechanical strain \cite{A77,A78,A79,A81,A84}. These works employed complementary straining techniques and provided a quantitative estimation of the strain level applied: bending of a flexible substrate \cite{A77,A78,A79}, using of a piezoelectric substrate \cite{A81}, and controllably wrinkling \cite{A84}. More recently, the use of substrates with very large thermal expansion coefficient have been used to engineer biaxial tensile strain in single-layer MoS$_2$ \cite{A83}. We refer the reader to previous section to find a detailed discussion about these different techniques to apply strain. 

\subsubsection{Homogeneous uniaxial strain}

\begin{figure}[t]
\includegraphics[scale=0.90,clip=]{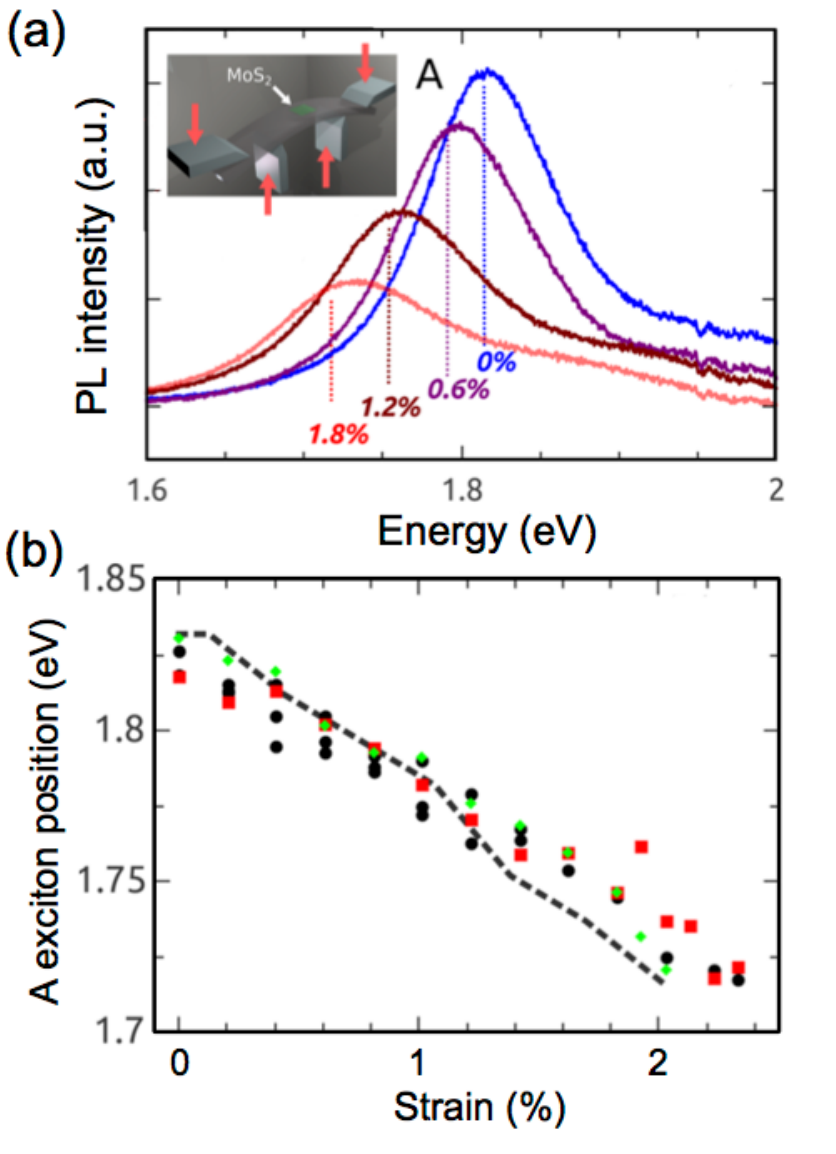}
\caption{
{\bf Band engineering with homogeneous uniaxial strain.}       (a) Photoluminescence spectra of a monolayer MoS$_2$ as it is strained from 0 to 1.8\%. (inset) Schematic of the beam bending apparatus used to strain MoS$_2$. (b) Strain dependent position of the A peak for several monolayers (colors represent different samples). Adapted from Ref. \cite{A77} with permission of the American Chemical Society.}
\label{Fig:Fig6}
\end{figure}

He {\it et al.}, Conley {\it et al.} and Zhu {\it et al.} transferred single- and bilayer MoS$_2$ onto flexible substrates and using a bending test experimental configuration they subjected these layers to homogeneous tensile uniaxial strain \cite{A77,A78,A79}. In these works it was found that the excitonic features, present in the photoluminescence emission and absorption spectra, linearly shift towards lower energy when the tensile strain is increased, indicating a reduction in the band gap upon homogeneous tensile uniaxial strain, as shown in Fig. \ref{Fig:Fig6}. The magnitude of the photoluminescence and/or absorption shifts upon strain reported in Refs. \cite{A77,A78,A79} are in relatively good agreement and discrepancies might be attributed to a difference in the PoissonÕs ratio of the different flexible substrates employed in the different works, as the bending of a beam induces not only an in-axis tension but also an off-axis compression whose magnitude is dictated by the substrate Poisson's ratio. Uniform uniaxial tensile strain has been also employed to tune the band structure of other 2D semiconductors beyond MoS$_2$. In Refs. \cite{A80,A88} a detailed analysis of the photoluminescence emission of WS$_2$ and WSe$_2$ nanosheets upon uniaxial tensile strain is reported.

\begin{figure}[t]
\includegraphics[scale=0.90,clip=]{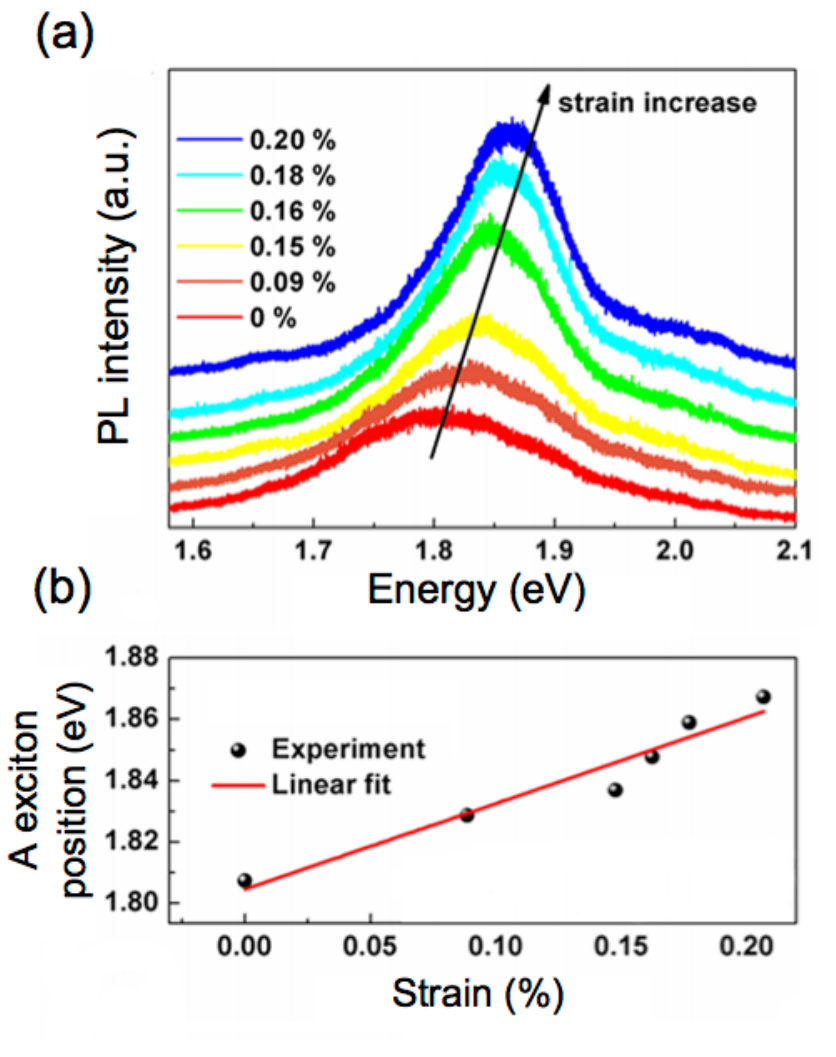}
\caption{
{\bf Band engineering with homogeneous biaxial strain. }    (a) Photoluminescence spectra of trilayer MoS$_2$ as it is strained with a piezoelectric substrate. (b) Strain dependent position of the A peak. Adapted from Ref. \cite{A81} with permission of the American Chemical Society.}
\label{Fig:Fig7}
\end{figure}

\subsubsection{Homogeneous biaxial strain}\label{Sec:Biaxial}

Hui {\it et al.} employed a piezoelectric substrate to apply homogeneous compressive strain to a trilayer MoS$_2$ flake, mechanically exfoliated onto a [Pb(Mg$_{1/3}$Nb$_{2/3}$)O$_3$]$_{0.7}$-[PbTiO$_3$]$_{0.3}$ substrate \cite{A81}. A CVD grown graphene layer was transferred on top of the trilayer MoS$_2$ flake as transparent and flexible top electrode to apply a uniform bias voltage to the piezoelectric material. The changes in the electronic band structure, induced by the straining with the piezoelectric substrate, were also monitored by the shift of the A excitonic feature in the photoluminescence spectra. Fig. \ref{Fig:Fig7}(a) shows a collection of photoluminescence spectra acquired at different compressive strain levels. The position of the photoluminescence peak, due to the generation of the A exciton, as a function of the strain is displayed in Fig. \ref{Fig:Fig7}(b). From this measurement Hui {\it et al.} determined a tunability of the A exciton of trilayer MoS$_2$ of $+300$ meV/\% of compressive biaxial strain. 

\begin{figure}[t]
\includegraphics[width=\columnwidth]{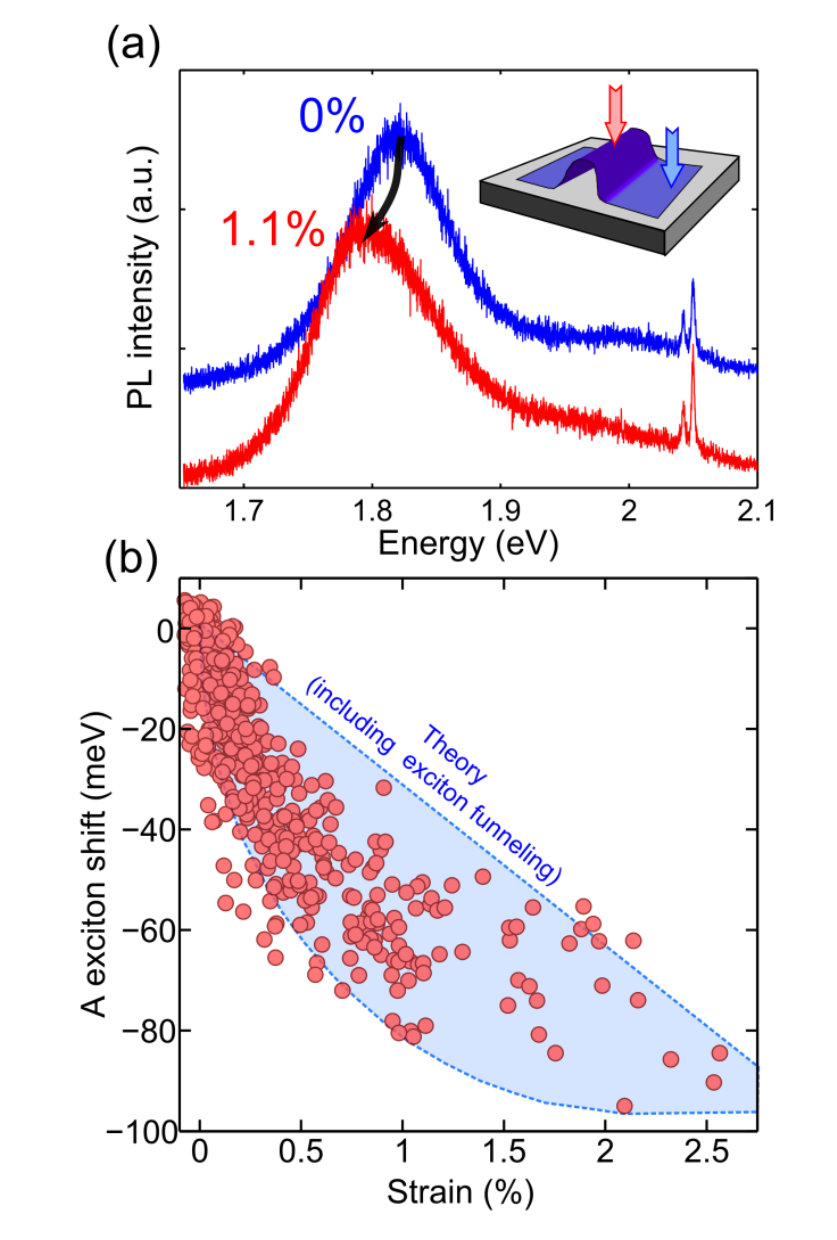}
\caption{
{\bf Band engineering with inhomogeneous uniaxial strain.}    (a) Photoluminescence spectra of a trilayer MoS$_2$ acquired at two different locations: flat (0\% strain) and on top of a wrinkle (1.1\% strain). (b) Strain dependent position of the A exciton peak. According to a tight binding calculation, including a correction to account for the exciton funnelling, the expected data lay within the light-blue area. Adapted from Ref. \cite{A84} with permission of the American Chemical Society.}
\label{Fig:Fig8}
\end{figure}

As discussed in the previous section, another method to apply homogeneous biaxial strain to 2D semiconductors is to transfer them to a substrate with a large thermal expansion coefficient and heating up the substrate. In this way the substrate will expand at a higher rate than the 2D semiconductor material and thus (if the adhesion to the substrate is strong enough to avoid slippage) a net biaxial tension will be applied to the 2D material. This straining technique has been employed by Plechinger {\it et al.} to strain monolayer MoS$_2$. By using a poly-dimethil siloxane substrate instead of conventional dielectric substrates large biaxial strain can be achieved \cite{A83}. Note that the thermal expansion coefficient of poly-dimethil siloxane is 1000 times larger than that of SiO$_2$ and 150 times larger than that of bulk MoS$_2$. By heating up the substrate to 150 \degree C a biaxial tensile strain of 0.23\% was achieved, yielding a shift of the A exciton of $-60$ meV. By measuring on a SiO$_2$ substrate (which has a thermal expansion smaller than MoS$_2$ and a poor adhesion) as a reference substrate Plechinger {\it et al.} determined that 40\% of the observed shift is due to the substrate induced biaxial tension and the other 60\% is due to the intrinsic expansion of MoS$_2$ upon heating. Therefore, a strain tunability of $-105$ meV/\% of tensile biaxial strain was found in these experiments. Similar results were also obtained by employing a focalized laser to increase the temperature of the substrate instead of a hotplate and the use of focalized laser heating to engineer more complex straining patterns was also explored in Ref. \cite{A83}.

\subsubsection{Non-homogeneous uniaxial strain}\label{Sec:Inhomogeneous1}

Alternatively to the previous experimental approaches to apply strain, that employed uniform straining techniques, Castellanos-Gomez {\it et al.} employed a controllable wrinkling method that make it possible to achieve highly localized strains \cite{A84}. Winkles are fabricated in the MoS$_2$ layers by transferring ultrathin MoS$_2$ flakes onto a pre-stretched poly-dimethil siloxane and subsequently releasing the strain (see the previous section for more details).

\begin{table*}
\begin{tabular}{|c|c|c|c|c|c|}
\hline
\hline
 & {\bf Work}  &{\bf Material}& {\bf Experiment} & {\bf Strain tunability}\\
\hline
\multirow{16}{2.4cm}{Homogeneous uniaxial}  & \multirow{3}{*}{Conley {\it et al.} \cite{A77}} &   1L MoS$_2$ & PL (A exciton) & $-45\pm 7$ meV/\% \\ \cline{3-5}
& & \multirow{2}{*}{2L MoS$_2$} & PL (A exciton) & $-53\pm 10$ meV/\% \\
& & & PL (I exciton) & $-129 \pm 20 $ meV/\% \\  \cline{2-5}
& \multirow{6}{*}{He {\it et al.} \cite{A78}} & \multirow{2}{*}{1L MoS$_2$} & PL \& Abs (A exciton) & $-64\pm 5$ meV/\% \\
& & & PL \& Abs (B exciton) & $-68\pm 5$ meV/\% \\ \cline{3-5}
& & \multirow{4}{*}{2L MoS$_2$} & PL (A exciton) & $-48\pm 5$ meV/\% \\
& & & Abs (A exciton) & $-71\pm 5$ meV/\% \\
& & & PL \& Abs (B exciton) & $-67 \pm 5$ meV /\% \\
& & & PL (I exciton) & $-77 \pm 5$ meV/\% \\ \cline{2-5}
 & \multirow{3}{*}{Zhu {\it et al.} \cite{A79}} &   1L MoS$_2$ & PL (A exciton) & $-48$ meV/\% \\ \cline{3-5}
& & \multirow{2}{*}{2L MoS$_2$} & PL (A exciton) & $-46$ meV/\% \\
& & & PL (I exciton) & $-86 $ meV/\% \\  \cline{2-5}
 & \multirow{3}{*}{Wang {\it et al.} \cite{A80}} &   \multirow{3}{*}{1L WS$_2$} & PL (A exciton) & $-11$ meV/\% \\ 
& & & PL (A$^-$ exciton) & $-11$ meV/\% \\
& & & PL (indirect) & $-19 $ meV/\% \\  \cline{2-5}
& Desai {\it et al.} \cite{A88} & 1L-4L WSe$_2$ & PL & N/A \\
\hline
\multirow{2}{2.4cm}{Homogeneous biaxial} & Hui {\it et al.} \cite{A81} & 3L MoS$_2$ & PL (A exciton) & $+300$ meV/\%$^*$ \\ \cline{3-5} 
 & Plechinger {\it et al.} \cite{A83} & 1L MoS$_2$ & PL (A exciton) & $-105$ meV/\%\\
 \hline
 \multirow{2}{2.4cm}{Inhomogeneous uniaxial} & Castellanos-Gomez {\it et al.} \cite{A84} & 3L-5L MoS$_2$ & PL (A exciton) & $-60$ meV/\% \\ \cline{2-5}
& Yang {\it et al.} \cite{A85} & 1L-2L ReSe$_2$ & PL (A exciton)&  $-37$ meV/\%\\ 
\hline
\end{tabular}
\caption{Comparison between the different strain engineering experimental results available in the literature for 2D semiconducting materials.}
\label{Tab:Works}
\end{table*}

\begin{figure}[t]
\includegraphics[scale=0.90,clip=]{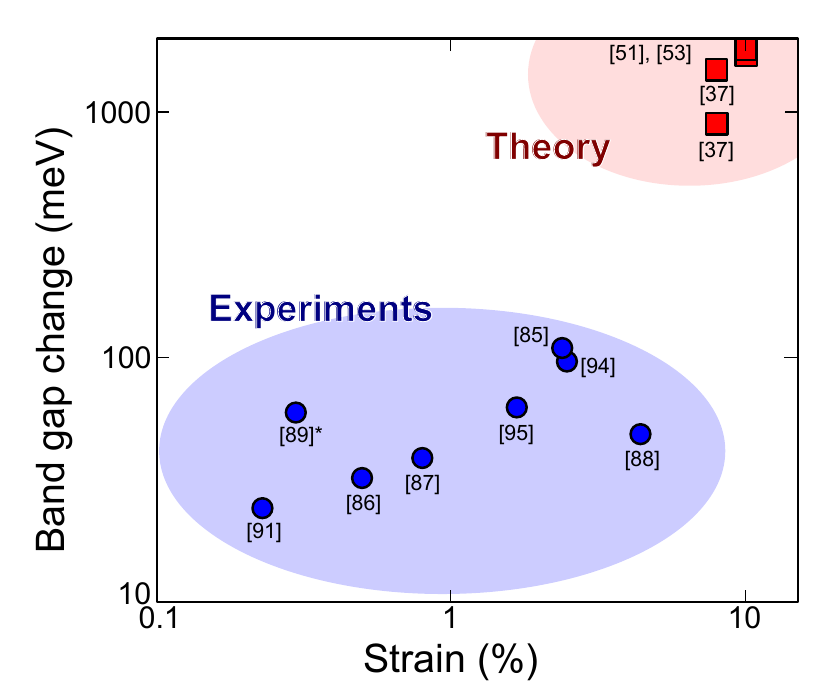}
\caption{
{\bf Summary of the state-of-the-art on experiments and theoretical predictions on strain engineering in atomically thin MoS$_2$, WSe$_2$ and ReSe$_2$.}     The figure also includes some theoretical predictions of the band gap change in single-layer MoS$_2$ upon very large strains for comparison. These works will be discussed in more detail in Sec. \ref{Sec:Theory}. }
\label{Fig:Fig9}
\end{figure}

The photoluminescence emission of the wrinkled MoS$_2$ samples depends on the position. On the flat regions of the sample (no strain) the photoluminescence shows a prominent peak corresponding to the generation of the A exciton (see Fig. \ref{Fig:Fig8}(a)). On the topmost part of the wrinkles (where the uniaxial strain due to the bending of the layer is maximum) the A exciton peak shifts towards lower energies as shown in Fig. \ref{Fig:Fig8}(a). This local variation of the photoluminescence indicates that the band structure spatially varies following the strain profile on the sample. This can be exploited to attract excitons, generated far from the maximum strain locations, towards the topmost part of the wrinkles \cite{A84}.  This concept, referred to in the literature as funneling effect, has been theoretically proposed by Feng {\it et al.} \cite{A45} and it will be discussed in more detail in the theory sections of this review article.

Fig. \ref{Fig:Fig8}(b) shows the strain dependence of the A-exciton shift, measured in several wrinkles with different maximum strain levels. Unlike the experimental data acquired with the previous approaches, the A-exciton shift shows a marked sublinear trend and a relatively large dispersion (larger than that expected from the measurement setup). A tight binding calculation shows that these two features can only be reproduced by including a correction in the theory that accounts for the drift of the photogenerated excitons towards regions with larger strain before recombination. Therefore, the presence of these two features can be considered the smoking gun of the funneling effect taking place in the inhomogeneously strained MoS$_2$ layer. 

The inhomogeneous uniaxial straining has been also applied to modify the bandstructure of ReSe$_2$ by Yang {\it et al.} \cite{A85}. In this work, the role of strain on the electrical and magnetic properties of ReSe$_2$ is also explored using the same wrinkling method.

\subsubsection{Summary of the different band engineering experiments}

In this section, we present a summary of the different strain engineering experiments carried out on 2D semiconducting materials. Table \ref{Tab:Works}, displays the strain tunability determined for different materials (MoS$_2$, WS$_2$, WSe$_2$ and ReSe$_2$) employing different straining techniques. Table \ref{Tab:Works} also displays the kind of experimental measurement (photoluminescence spectroscopy or absorption spectroscopy) as well as the spectroscopic feature studied (A, A$^-$, B or I exciton).

\begin{figure*}[t]
\includegraphics[scale=0.65,clip=]{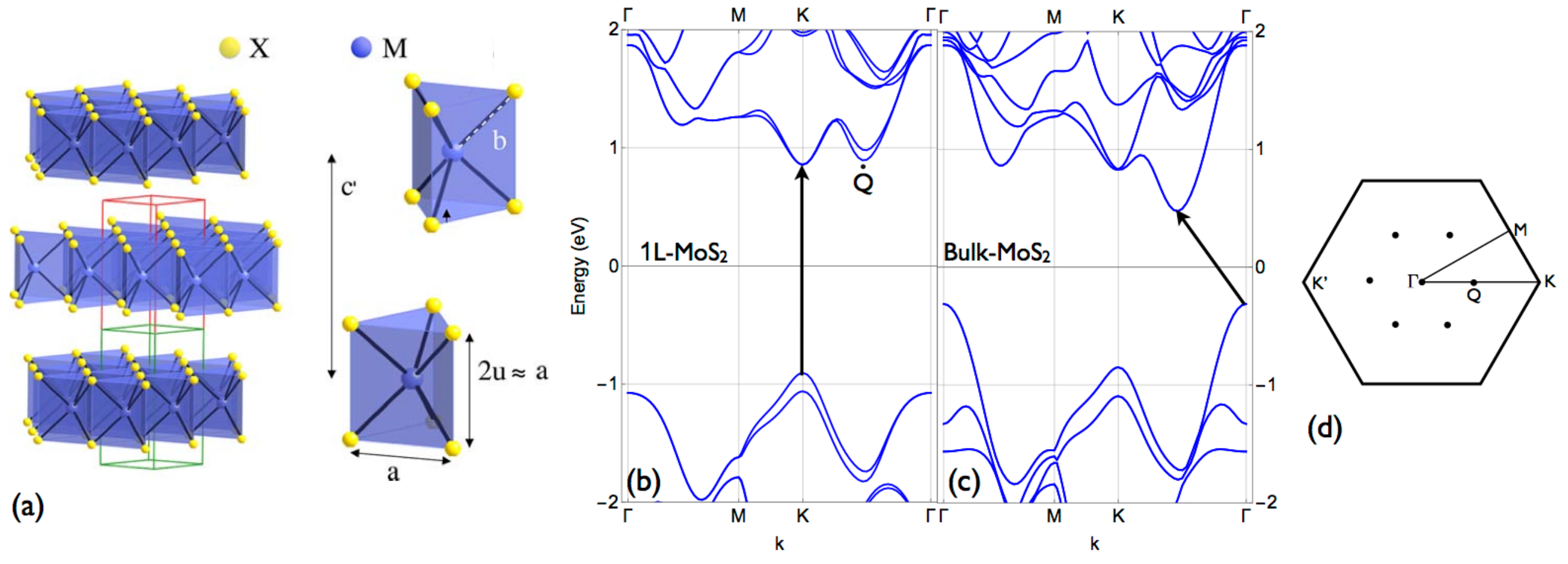}
\caption{
{\bf Lattice and electronic band structure of semiconducting TMDs.} (a) Crystal structure of TMD where blue spheres represent a transition metal atom and yellow spheres, a chalcogen element. (b) DFT electronic band structure of single-layer MoS$_2$ and (c) bulk MoS$_2$. Arrows indicate the gap in each case. (d) Two dimensional BZ.
}
\label{Fig:MX2}
\end{figure*}

Fig. \ref{Fig:Fig9} shows a graphic summary of the state-of-the-art on experiments on strain engineering in atomically thin 2D semiconductors. The Figure displays the maximum band gap change (measured with optical spectroscopy techniques) versus the maximum strain load achieved in each experimental work. Note that, in Fig. \ref{Fig:Fig9} experimental results obtained for different materials, with different straining techniques and applying different kind of strains are all plotted together. Some theoretical predictions of the band gap change for larger strain loads are also included in the Figure to illustrate that the field of strain engineering in atomically thin semiconducting materials is very young as the current experimental works have achieved moderate strains in comparison to the maximum strains that 2D semiconductors can stand before rupture. At those larger values of strain, one would expect a strain-induced tunability of the band gap about 10 times larger than the maximum tunability experimentally observed so far.

\section{Theoretical study of strain in 2D crystals }\label{Sec:Theory}

In this section we review the effect of strain on semiconducting 2D crystals from a theoretical point of view. We first describe the main features of the crystal and electronic structure of the most commonly studied 2D crystals in the absence of external strain. Then we discuss the effect of homogeneous and inhomogeneous strain on the optical and electronic properties. As we have seen in Sec. \ref{Sec:Experiment}, most of the current experimental results on strain engineering are performed on transition metal dichalcogenides, and for strain levels that do not exceed $\sim 3\%$. Here we include a revision of theoretical works on the field that make predictions for higher strain values than currently achieved experimentally, as well as for other families of 2D semiconducting crystals, like single and multi-layer black phosphorus, silicene and germanene, unexplored experimentally so far.

\subsection{Electronic structure}\label{Sec:NoStrain}

\subsubsection{Transition metal dichalcogenides}\label{Sec:TMDsNoStrain}

The most studied two dimensional crystals different from graphene are the semiconducting transition metal dichalcogenides, $MX_2$ (where $M = $Mo, W and $X$ = S, Se, Te). The crystal structure of monolayer and multilayer  {\em MX}$_2$
is schematically shown in Fig.  \ref{Fig:MX2}(a). {\em MX}$_2$ is composed of an inner layer of metal $M$ atoms ordered
in a triangular lattice sandwiched between two layers
of chalcogen $X$ atoms lying on the triangular net of alternating
hollow sites \cite{KZH11,LX13,KGF13,KA12,RO14,RCG13}. In Table \ref{Tab:LattParam} we give the values of the $MX_2$ lattice parameters.
Similarly to graphene, the in-plane Brillouin zone (BZ) is an hexagon characterized by the
high-symmetry points $\Gamma=(0,0)$, K$=4\pi/3a(1,0)$,
and M$=4\pi/3a (0,\sqrt{3}/2)$, where $a$ is the in-plane $M-M$ or $X-X$ distance.

\begin{table}
\begin{tabular}{|l|c|c|c|}
\hline
\hline
 & $a$ & $u$ & $c'$ \\
\hline
MoS$_2$ & $3.160$ & $1.586$ & $6.140$ \\
\hline
WS$_2$ & $3.153$ & $1.571$ & $6.160$ \\
\hline
MoSe$_2$ & $3.288$ & $1.664$ & $6.451$ \\
\hline
WSe$_2$ & $3.260$ & $1.657$ & $6.422$ \\
\hline
\hline
\end{tabular}
\caption{
Lattice parameters (in \AA\, units) of TMDs, as given in Ref. \cite{BMY72,SBJ87,KA12}. $a$ is the $M-M$ atomic distance, $c'$ is the distance between the $M$ layers, and $u$ the internal vertical distance between the $M$ 
and the $X$ planes.
For bulk compounds, the $z$-axis crystal parameter is given by $c=2c'$.
}
\label{Tab:LattParam}
\end{table}

 \begin{figure*}[t]
\includegraphics[scale=0.65,clip=]{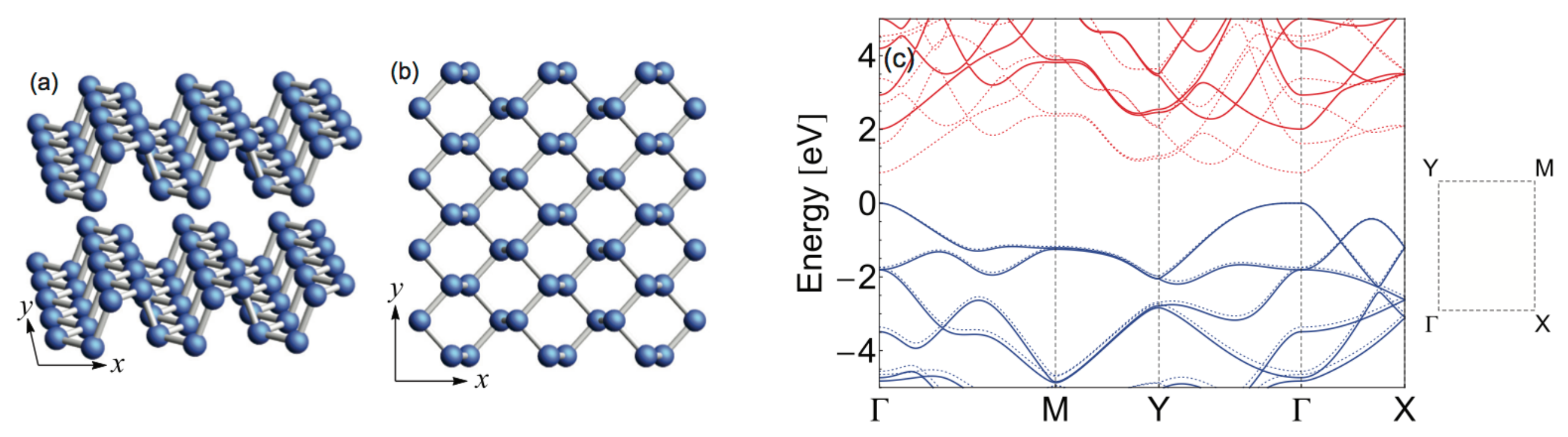}
\caption{
{\bf Lattice and electronic band structure of black phosphorus.} (a) Crystal structure of BP where blue spheres represent P atoms. (b) Top view of single-layer phosphorene. (c) DFT (dashed lines) and $GW$ (solid lines) electronic band structure of phosphorene. The rectangular BZ is also shown. Adapted from Ref. \cite{TY14} with permission of the American Physical Society.
}
\label{Fig:BP}
\end{figure*}

The orbital character of the electronic bands of TMDs is studied in detail in Ref. \onlinecite{CG13}. The conduction and valence bands are made by hybridization of the $d_{xy}$, $d_{x^2-y^2}$ and $d_{3z^2-r^2}$ orbitals of $M$, and the $p_x$, $p_y$ and $p_z$ orbitals of $X$. All the single-layer $MX_2$ compounds are direct gap semiconductors, with the gap placed at the two inequivalent K and K' points of the BZ, as seen in Fig. \ref{Fig:MX2}(b) for MoS$_2$. The most relevant contribution at the edge of the valence band is due to a mixture of $d_{xy}$ and $d_{x^2-y^2}$ orbitals of the metal $M$, hybridizing with the $p_x$ and $p_y$ orbitals of the $X$ chalcogen atoms. The main contribution to the edge of the conduction band is due to $d_{3z^2-r^2}$ of the metal $M$, plus a minor contribution of $p_x$ and $p_y$ orbitals of the chalcogen $X$ \cite{CG13}. In Table \ref{Tab:Gaps}  we give the energy gap values of the different compounds discussed in this review. Multi-layers and bulk TMD compounds are indirect gap semiconductors, with the maximum of the valence band at $\Gamma$ point and the minimum of the conduction band at the Q point, which is situated midway between $\Gamma$ and K points of the BZ.

\begin{table}
\begin{tabular}{lccc|c|}
\hline
\hline
  &\multicolumn{3}{c}{Band Gap}\\
\cline{2-5}
 & Monolayer & Bilayer & Bulk & Ref.   \\
\cline{2-5}
MoS$_2$  & 1.715 & $1.710-1.198$ &  $1.679-0.788$ &  \cite{RO14}\\
WS$_2$   & 1.659 & $1.658-1.338$ & $1.636-0.917$ &  \cite{RO14}   \\
MoSe$_2$ & 1.413 & $1.424-1.194$ & $1.393-0.852$ &  \cite{RO14}     \\
WSe$_2$  & 1.444 & $1.442-1.299$ & $1.407-0.910$ &  \cite{RO14}   \\
 ReSe$_2$  & 1.43 & $ - $ & $ 1.35 $ &  \cite{Tongay14}   \\
BP & 1.60 & 1.01 & 0.10 &  \cite{RK14}\\
Silicene & $1.6\times 10^{-3}$ & -- & -- &  \cite{LFY11}\\
Germanene & $24\times 10^{-3}$ & -- & -- &  \cite{LFY11}\\
\hline
\hline
\end{tabular}
\caption{Band gap of 2D crystals, expressed in eV, as obtained from first principle calculations. Notice that all monolayer samples for MoS$_2$, WS$_2$, MoSe$_2$ and WSe$_2$ present a direct gap. The two values given for bilayer and bulk samples for those compounds correspond to the sizes of the direct/indirect gaps, respectively. The gap of ReSe$_2$ is direct for both, bulk and single layer. Results for BP are taken from $GW$ calculations \cite{RK14}. Results for silicene and germanene are taken from relativistic DFT calculations, and corresponds to the low-buckled configuration (see text and Ref. \onlinecite{LFY11}). }
\label{Tab:Gaps}
\end{table}

One of the main characteristics of TMDs is the strong spin-orbit coupling (SOC), which leads to a splitting of the valence and conduction bands. The valence band of MoS$_2$ and MoSe$_2$ is split by $\sim$150 meV, whereas the splitting for WS$_2$ and WSe$_2$ is considerably larger, of the order of $\sim$ 400 meV, due to the heavier mass of W. The conduction band present a SOC splitting of several tens of meV at the band edge at the K point \cite{KGF13}, as well as at the Q point \cite{RO14-2D}, as shown in Fig. \ref{Fig:MX2}(b). The phonon spectrum of TMDs is also very rich, and we refer the reader to Ref. \cite{LS14b} for a comprehensive review on the subject.

\subsubsection{Black Phosphorus}

\begin{figure*}[t]
\includegraphics[scale=0.65,clip=]{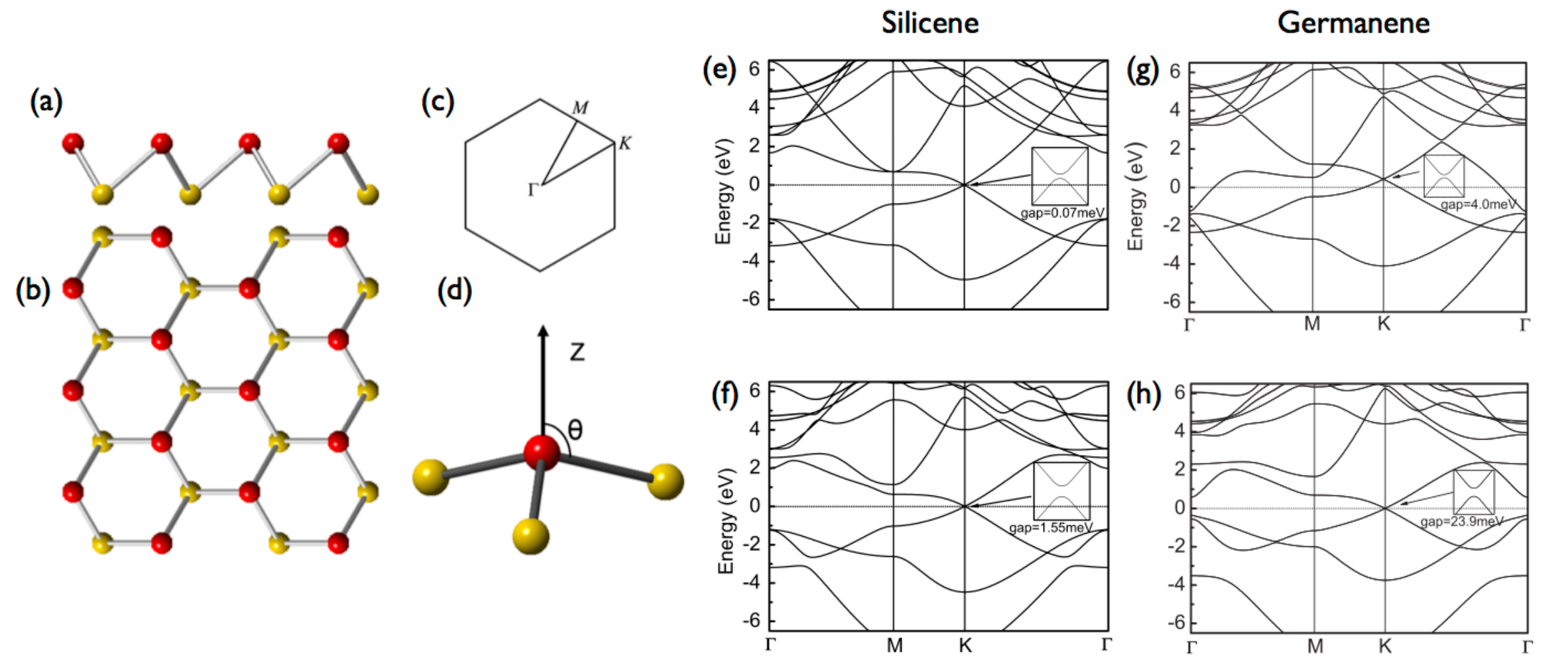}
\caption{
{\bf Lattice and electronic band structure of silicene and germanene.} (a) Side view of lattice geometry of low-buckled silicene and germanene. (b) Top view of the crystal structure. (c) First Brillouin zone of silicene and germanene. (d) Angle that defines the Si-Si bond and the $z$ direction normal to the layer. (e) and (f) show the relativistic band structures of silicene for $\theta=90\degree$ and $\theta=102\degree$, respectively. The insets show a zoom near the K point of the BZ of the energy dispersion, with the gap induced by SOC. (g) and (f) are the same than (e) and (f), but for germanene. Adapted from Ref. \cite{LFY11} with permission of the American Physical Society.
}
\label{Fig:Silicene}
\end{figure*}

Bulk black phosphorus (BP) is another layered van der Waals crystal consisting of puckered atomic sheets of phosphorus weakly coupled together \cite{M86}. The electronic and elastic properties of BP have been also studied in its nanotube structure \cite{SH00}.  A single layer of BP consists of a corrugated arrangement of P atoms with a thickness of $\sim5$ \AA~[see Fig. \ref{Fig:BP}(a)-(b)]. Bulk BP is obtained by alternate stacking of the layers along the [001] direction. The BP intra-layer bonding is due to the $sp^3$ hybridization of P atoms.

The band structure of single-layer BP is shown in Fig. \ref{Fig:BP}(c) as calculated from DFT and and $GW$ methods. The orbital contribution to the valence and conduction bands close to the gap has been studied in Ref. \cite{RK14}. For the valence band, the orbital composition at the $\Gamma$ point can be written as $|\Psi^{\rm v}(\Gamma)\rangle=0.17|s\rangle+0.40|p_x\rangle+0.90|p_z\rangle$ whereas for the conduction band we have $|\Psi^{\rm c}(\Gamma)\rangle=0.57|s\rangle+0.44|p_x\rangle+0.69|p_z\rangle$. The $p_z$ orbital has the highest contribution in both bands, but the $s$ and $p_x$ characters are not negligible. Interestingly, the $p_y$ orbital does not contribute to the formation of the valence and conduction bands. 

The gap at the $\Gamma$ point obtained from $GW$ calculations is $\sim 1.60$~eV for single-layer BP \cite{RK14}, which agrees well with photoluminescence measurements \cite{CZ14}. As we increases the number of layers, the gap decreases monotonically reaching the value of $\sim 0.1$~eV for the bulk material \cite{RK14} (see Table \ref{Tab:Gaps}). The in-plane electronic and optical properties of black phosphorus are highly anisotropic, due to the characteristic puckered structure of this crystal \cite{XWJ14,LG14}.

\subsubsection{Silicene and Germanene}

Silicene is another 2D crystal that is attracting high interest. Encapsulated silicene field effect transistor have been fabricated recently, showing Dirac-like ambipolar transport \cite{TA15}. This material consists of a honeycomb lattice of Si atoms, similar to graphene, as shown in Fig. \ref{Fig:Silicene}. However, the two triangular sublattices made by the Si  atoms are vertically displaced by 0.46~\AA, forming a buckled structure that resembles that of phosphorene.  The interatomic distance of silicene is 2.28~\AA, larger than that of graphene due to the large ionic radius of Si as compared to C. One can characterize the buckling of the 2D lattice by means of the angle $\theta$ between the Si-Si bond and the direction normal to the plane, as sketched in Fig. \ref{Fig:Silicene}(d). Different configurations can be defined, like the $sp^2$ (planar) for $\theta=90\degree$, the low-buckled which corresponds to $\theta=101.73\degree$, and the $sp^3$ configuration, that corresponds to $\theta=109.47\degree$ \cite{LFY11}. Due to the buckling of the lattice, the carriers move in a hybridization of the $p_z$ orbitals with the $\sigma$ orbitals, leading to a large SOC as compared to graphene.  As a consequence, an appreciable energy gap can be opened at the Dirac points (see Table \ref{Tab:Gaps}). 

Although it has not yet synthesized in experiments, {\it germanene}, a single layer of Ge atoms, is another 2D crystal that it is attracting a lot of interest. The crystal structure is similar to that of silicene, with Ge-Ge distance of $2.42$~\AA. The estimated gap, based on relativistic DFT calculations, is also higher than in silicene (see Fig. \ref{Fig:Silicene}(g)-(h)). In both cases, hydrostatic strain is expected to increase the size of the gap, what can help for the observation of quantum spin Hall effect in the low-buckled structure of these compounds \cite{LFY11}.

\begin{figure*}[t]
\includegraphics[scale=0.60,clip=]{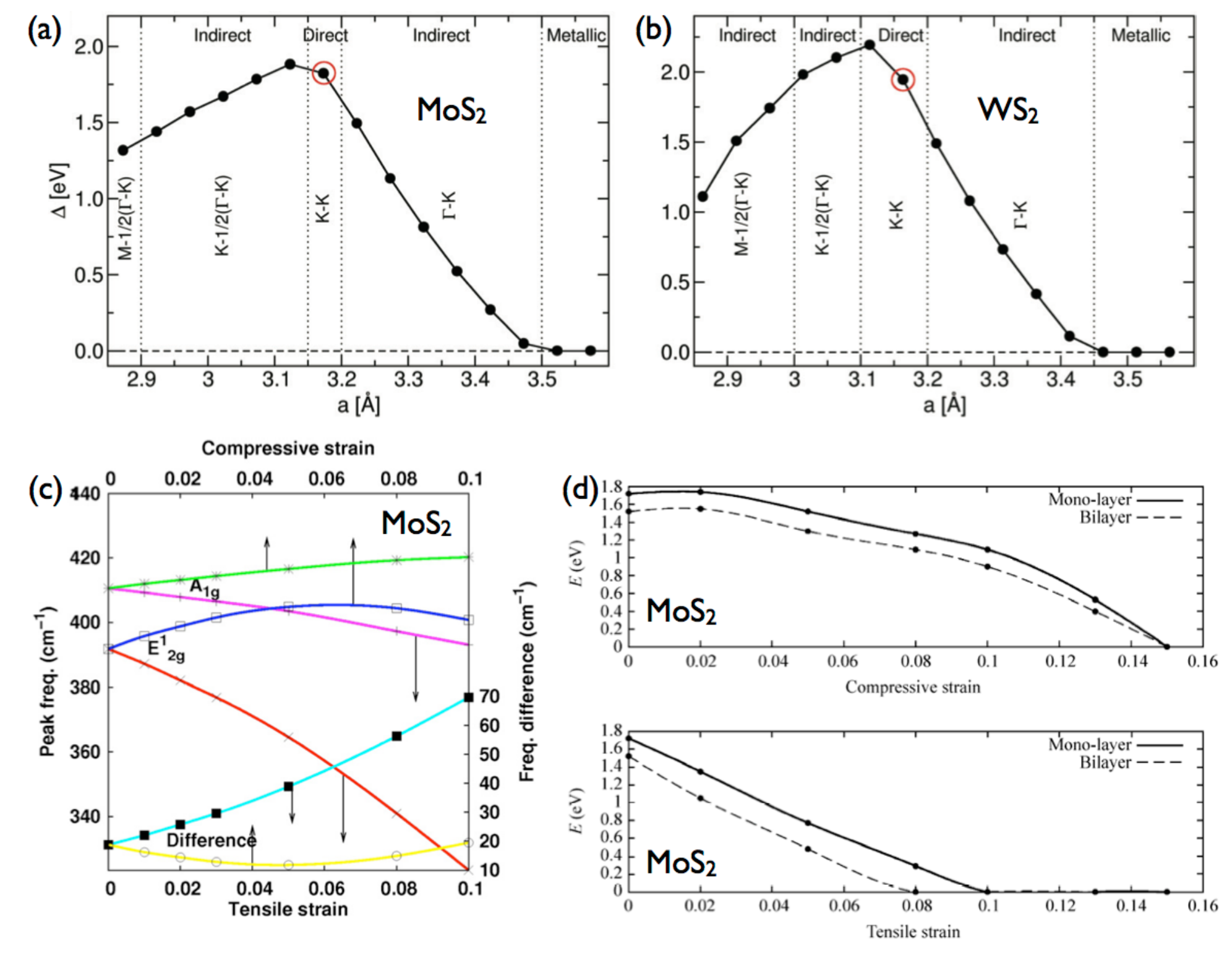}
\caption{
{\bf Homogeneous strain in TMDs.} Evolution of the band gap of single layer MoS$_2$ (a) and WS$_2$ (b) under uniform isotropic tensile and compressive strain.  The edges of the valence and conduction bands evolve with strain, and the points of the BZ at which they occur are indicated. The red circles mark the equilibrium structures in the absence of strain \cite{A46}. (c) Dependence of the phonon frequency of the $E^1_{2g}$ and $A_{1g}$ Raman modes, as well as their difference, versus applied strain (tensile: red and magenta; compressive: green and blue; difference: yellow and cyan) \cite{SS12}. (d) Evolution of the band gap with biaxial compressive (top) and tensile (bottom) strain for single-layer and bilayer MoS$_2$ \cite{A44}. Panels (a) and (b) adapted from Ref. \cite{A46} with permission of the American Physical Society. Panel (c) adapted from Ref. \cite{SS12} with permission of Springer. Panel (d) adapted from Ref. \cite{A44} with permission of Elsevier.
}
\label{Fig:HomogTMDs}
\end{figure*}

\subsection{Homogeneous strain}\label{Sec:Homogeneous}

\subsubsection{Transition metal dichalcogenides}

The effect of homogeneous strain in the band structure of 2D semiconducting crystals has been the focus of many theoretical papers. For TMDs, a number of DFT calculations have studied the dependence of the band structure and phonon modes with external strain \cite{SS12,A44,YL12,JS12,PZ12,A46,shi13,zhang2013,guzman,jiang,Horzum2013},
and the consequences on different physical properties have been discussed in
Refs. \cite{bhatta,hosseini,rhim}.
 In Fig. \ref{Fig:HomogTMDs} it is shown the evolution of the band gap with uniform isotropic strain for single layer MoS$_2$ (a) and WS$_2$ (b). In the absence of strain, TMDs are direct gap semiconductors, as discussed in Sec. \ref{Sec:TMDsNoStrain}, with the gap placed at the K point of the BZ. Uniform (uniaxial or biaxial) tensile strain leads to a linear decrease of the band gap, and a direct-to-indirect gap transition is reached, with the valence band edge being now situated at the
$\Gamma$ point \cite{wangkutana}. Further tensile strain eventually closes the gap, leading to a semiconducting-to-metal transition for elongations of the order of 11\% \cite{SS12,SS12b,A46}. This transition is expected to be possible in the lab, since for such strain the metal-chalcogen bonds are stretched, but not yet broken \cite{A46}. The case of compressive strain is similar, although its influence in the band structure is smaller as compared to tensile strain, as it can be seen in the top panel of Fig. \ref{Fig:HomogTMDs}(d). Notice that, due to the relatively small bending rigidity, compressive strain is difficult to be applied to suspended TMDs samples, although substrate-induced compression is possible by depositing layers of TMDs on an elastomeric substrate, as discussed in Sec. \ref{Sec:Elongatation}.

\begin{figure*}[t]
\includegraphics[scale=0.60,clip=]{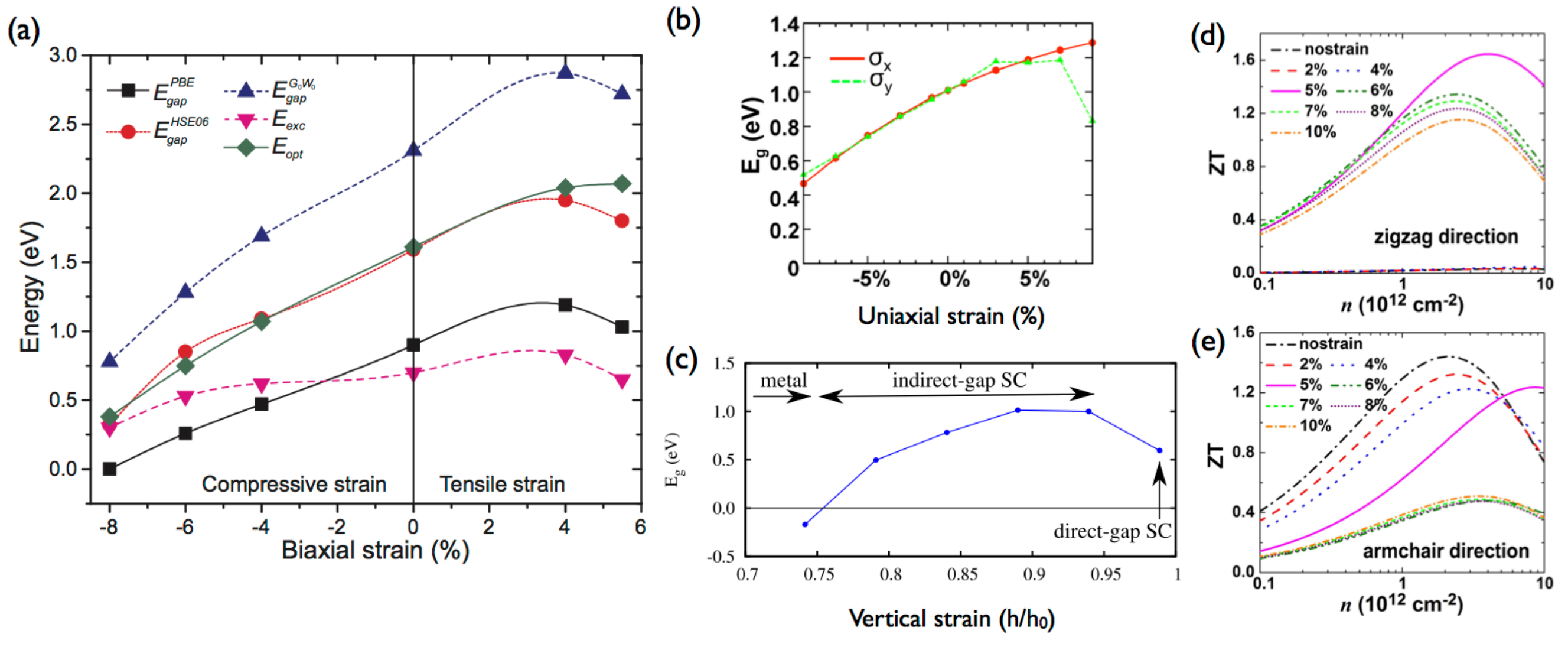}
\caption{
{\bf Homogeneous strain in BP.} (a) Evolution of the electronic ($E_{gap}$) and optical ($E_{opt}$) band gaps of single layer BP with biaxial strain. The dependency of the excitonic binding energy $(E_{exc})$ is also shown. The different numerical methods used are indicated as superscripts \cite{CSP14}. (b) Band gap evolution with uniaxial strain \cite{LY14}.  (c) Band gap with vertical compressive strain. The thickness of the unstrained layer is $2h_0$ \cite{RCC14}. (d)-(e) Dimensionless figure of merit $ZT$ as a function of carrier density for different values of uniaxial strain applied along the zigzag (d) and armchair (e) directions. Panel (a) adapted from Ref. \cite{CSP14}, panel (c) adapted from Ref. \cite{RCC14} and panels (d) and (e) adapted from Ref. \cite{LS14}, with permission of the American Physical Society. Panel (b)  adapted with permission from Ref. \cite{LY14}, copyright 2014 American Chemical Society.
}
\label{Fig:HomogBP}
\end{figure*}

Application of uniaxial tensile strain, on the other hand, is possible in the laboratory by the use of different techniques, as reviewed in Sec. \ref{Sec:Techniques}. DFT calculations for this kind of strain result in a lowering of the size of the band gap, as in the uniform case, with a possible transition to a metallic state. The electron mass at the Q point of the conduction band is found to increase (decrease) with uniaxial tensile (compressive) strain, while the effective masses at K decrease (increase) with increasing uniaxial tensile (compressive) strain \cite{PV12}. The effect of mechanical deformations on the electronic band structure is especially relevant at the K point of the BZ, where both edges of the conduction and valence bands are significantly lowered as compared to the unstrained case \cite{A46,PV12}. On the other hand, the top of the valence band at $\Gamma$ moves upwards. This is due to the fact that the orbital overlap between the metal and chalcogen atoms decreases due to the change in the bond distances with applied strain \cite{A46}.

\subsubsection{Black Phosphorus}

The effect of strain on the electronic and mechanical properties of single layer black phosphorus have been studied using first-principle methods \cite{AS12,RCC14,LC14,GZT14,FL14,WP14,PWC14,LY14,EA15}. A detailed comparison of the BP electronic band structure using different numerical methods has been presented in Ref. \onlinecite{CSP14}. Inclusion of many-body effects within $G_0W_0$ increases the band gap as obtained from GGA-PBE. It was found that the electronic and the optical gaps, as well as the exciton binding energy increases (decreases) with the application of tensile (compressive) strain \cite{PWC14,CSP14}. This trend is shown in Fig. \ref{Fig:HomogBP}(a). The effect of uniaxial strain is similar, as shown in Fig. \ref{Fig:HomogBP}(b), with a direct-to-indirect gap transition predicted for compressive strain of the order of $\sim 5\%$ \cite{LY14}.  Vertical strain applied perpendicular to the BP layer has been studied in Ref. \onlinecite{RCC14}. This kind of strain leads first to a direct-to-indirect band gap transition, and for larger strains also a semiconducting-to-metal transition is predicted for uniaxial stress of the order of 24 GPa.

BP has been proposed as an good material for thermoelectric applications. Usually the dimensionless {\it figure of merit} $ZT=\sigma TS^2/(\kappa_e+\kappa_p)$ quantifies the performance of a thermoelectric material, where $\sigma$ is the electrical conductivity,  $T$ is the temperature, $S$ is the Seebeck coefficient, $\kappa_e$ is the lattice thermal conductivity and $\kappa_p$ is the electronic thermal conductivity. Crystals with a large Seebeck coefficient and electrical conductivity and/or low thermal conductivity are expected to behave as good thermoelectric materials. Uniaxial strain has been proposed as a route to enhance the Seebeck coefficient and the electrical conductivity in phosphorene, suggesting strain engineering as an effective method to increase the thermoelectric performance of BP \cite{LS14,LC14}.

First principle calculations have found that biaxial strain enhances significantly the electron-phonon coupling in black phosphorus. This effect has been proposed to favor superconductivity, leading to an increase of the superconducting critical temperature from 3 K to 16 K at typical carrier densities of $\sim 3.0 \times 10^{14}~ {\rm cm}^{-2}$. This phenomenon   is attributed to the simultaneous increase of the density of states around the Fermi level and the softening of the phonon modes. Furthermore, the anisotropy of the phosphorene lattice leads to different effects on the electronÐphonon coupling under uniaxial strain applied in different directions, leading to a $T_c\approx 10$~K and 8 K for uniaxial strain applied along the $x$- and $y$-directions, respectively \cite{GY15}.

\subsubsection{Silicene and germanene}

\begin{figure}[t]
\includegraphics[scale=0.28,clip=]{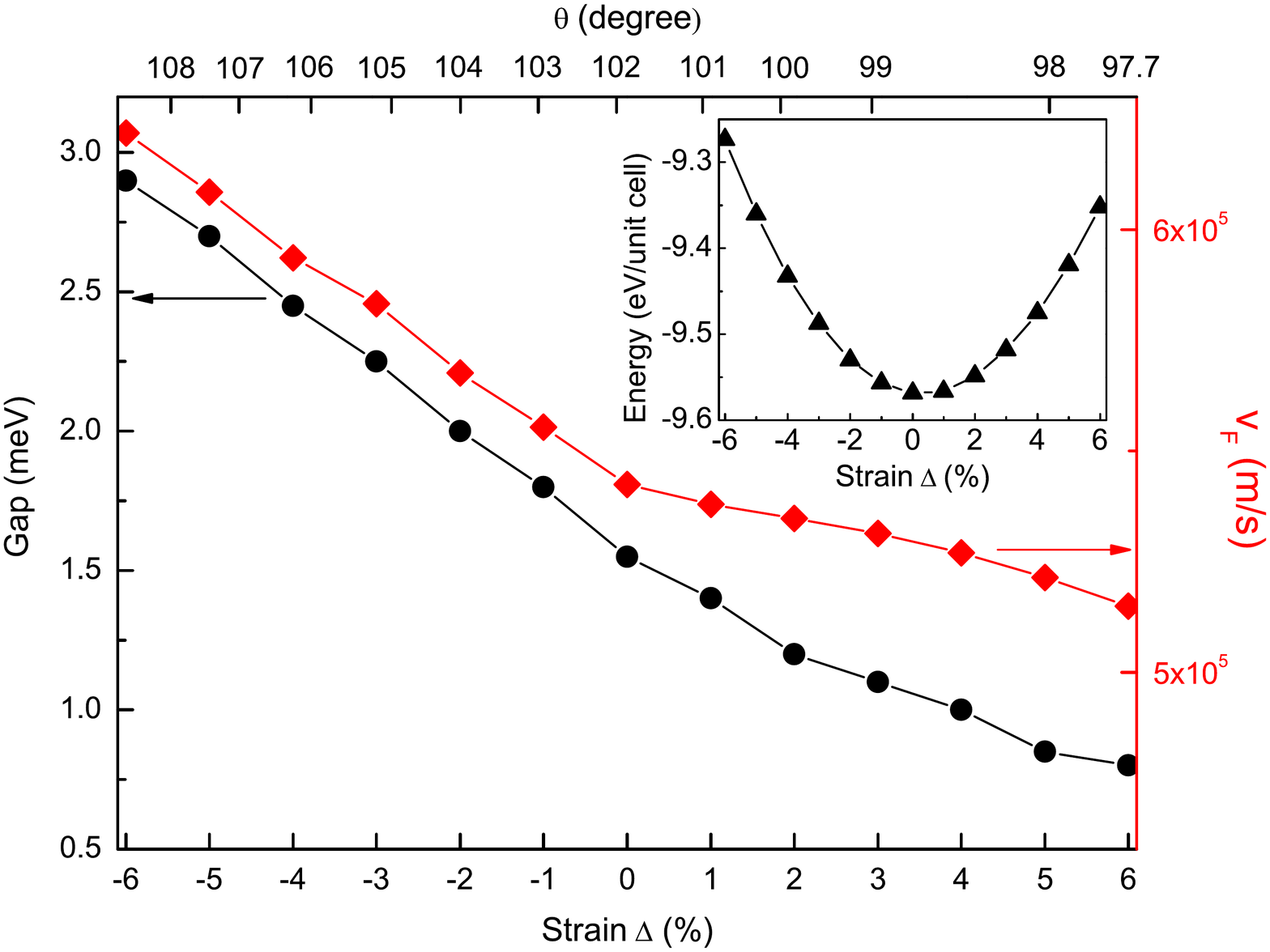}
\caption{
{\bf Homogeneous strain in silicene.} Dependence of the electronic energy gap (black circles) and Fermi velocity $v_F$ (red diamonds) near the Dirac points calculated from relativistic first-principles method, under different applied strain. The inset shows the energy of unit cell for different hydrostatic strain. Adapted from Ref. \cite{LFY11} with permission of the American Physical Society.
}
\label{Fig:HomogSilicene}
\end{figure}

The band structures of silicene and germanene are also affected by homogeneous strain \cite{LFY11,CC13,KS13}. The results show that the size of the gap at Dirac points grows with compressive strain, and decreases with tensile strain, as shown in Fig. \ref{Fig:HomogSilicene}.  It is also observed that the gap is higher for higher angle $\theta$, as defined in Fig. \ref{Fig:Silicene}(d). On the other hand, the Fermi velocity is only weakly affected by strain, being its value slightly less than the typical $v_F$ for graphene, this is $\sim 10^6{\rm m/s}$, due to the larger Si-Si interatomic distance. 

The characteristic buckled structure of silicene and germanene leads to a strain-induced self-doping phenomenon, as predicted by first-principle calculations of Ref. \cite{WD13}. The Dirac point is found to move below the Fermi level for compressive strain, leading to $n-$doped samples, whereas $p-$type doping can be achieved under tensile strain.

\subsection{Inhomogeneous strain}\label{Sec:Inhomogeneous}

\begin{figure*}[t]
\includegraphics[scale=0.65,clip=]{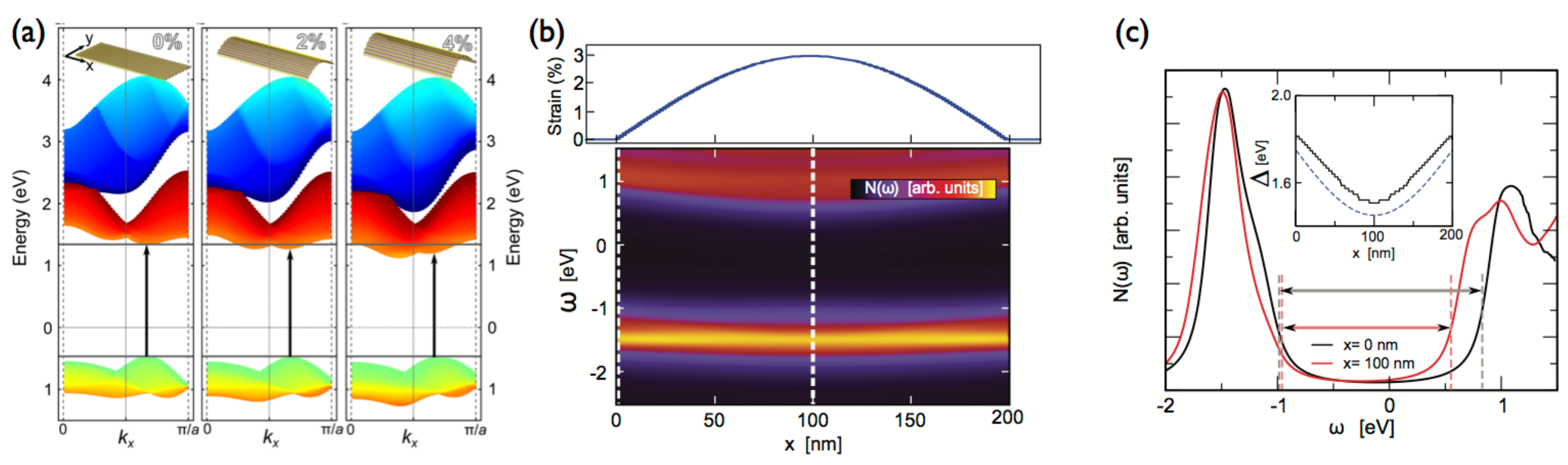}
\caption{
{\bf Inhomogeneous strain in TMDs.} (a) Band structure for a MoS$_2$ ribbon with zigzag edges. Periodic boundary conditions have been used, and a non-uniform strain profile as given by Eq. (\ref{Eq:StrainNonUniform}), where $\varepsilon_{\rm max}$ correspond to 0\%, 2\%  and 4\%, respectively. The arrows indicate the direct transitions at the K point of the BZ, which are observed by photoluminescence experiments. (b) Color map of the local density of states of a 200 nm MoS$_2$ wrinkle, as a function of the position along the wrinkle. The strain profile is sketched in the top panel. (c) LDOS calculated at the center $x=100$~nm (maximum strain) and at the edge $x=0$ (unstrained). The arrows denote the estimated band edges as given by the second derivative method.  Adapted with permission from Ref. \onlinecite{A84}, copyright 2013 American Chemical Society.
}
\label{Fig:InhomogTMDs}
\end{figure*}

In the previous section we have seen how homogeneous strain can be used to control the electronics and optical properties of 2D semiconductors. In this section we will see how strain engineering can be used to modify and control those properties locally, by means  of application of non-uniform strain. The possibility to continuously control the energy band gap in semiconductors  is of high interest for their application in optoelectronics devices. In particular, an inhomogeneous strain field acting on a 2D semiconducting crystal leads to a continuous spatial variation of its electronic band structure \cite{A84}. This effect has been proposed for the design of solar energy funnel, in which a broad band of the spectrum can be absorbed. As discussed in Sec. \ref{Sec:Inhomogeneous1}, this possibility was proposed theoretically by Feng {\it et al.} \cite{FL12} and demonstrated experimentally by Castellanos-Gomez {\it et al.} in Ref. \onlinecite{A84} for MoS$_2$. 

A different route to achieve an exciton funnel has been proposed in Ref. \cite{WQL14}, where it was shown, based on first principle calculations, that a small twist between two semiconducting atomic layers creates
an internal stacking translation $u({\bf r})$ that varies with position $\bf r$ and controls the local electronic bandgap $E_g(u({\bf r}))$. The authors of Ref. \cite{WQL14} propose the use of  carbon/boron
nitride or black phosphorus bilayer, and predict an energy gap with multiple local minima (which can act as funnel centers). The possibility to locally tune the band gap in non-planar single layer black phosphorus has been studied theoretically in Ref. \cite{Mehboudi15}. In that paper it was found that the gap can be reduced by 20\% for non-planar phosphorene allotropes, with respect to its value on planar crystals. Triaxial in-plane strain has been proposed to engineer the gap in h-BN monolayer \cite{Neek2013}.

\subsubsection{Tight-binding approach considering inhomogeneous strain: application to TMDs}

Theoretically, the study of non-uniform strain by means of first-principle methods is extremely expensive in computational terms, due to the huge unit cell that must be considered. An affordable alternative is the use of tight-binding methods \cite{SK54}, which are a good base to further include electron many-body effects, electron-lattice interaction, etc. Using geometric arguments,
the energy hopping integral $V_{l,l'}$ between two orbitals can be written in terms
of Slater-Koster two-center parameters $V_{ss\sigma}$, $V_{sp\sigma}$, $V_{pp\sigma}$, $V_{pp\pi}$, $V_{dd\sigma}$, $V_{dd\pi}$, $V_{dd\delta}$, $V_{pd\sigma}$, $V_{pd\pi}$, etc., which, together with the crystal fields, represent the
suitable tight-binding parameters to be optimized \cite{CG13}.

 Within the Slater-Koster framework, hopping
processes between two atoms depend only on the relative angle and on the relative atomic separation $R_{ij\mu\nu}=|{\bf R}_{ij\mu\nu}|$, through the hopping integrals
$V_{i,j,\mu,\nu}(R_{ij\mu\nu})$, where $i,j$ label the unit cell and the
$\mu,\nu$ indices denote the specific atom  and orbital within the
unit cell.
When considering the effects of homogeneous and inhomogeneous strain it is common to assume,
following Refs. \cite{CG09,VKG10}, that the main contribution comes from the
modulation of the hopping integrals $V_{i,j,\mu,\nu}(R_{ij\mu\nu})$
on the distance, neglecting the weak dependence on
the relative angles.

For a given profile of the strain tensor $\hat{\varepsilon}(x,y,z)$,
assuming the characteristic length of the sample to be
much larger than the lattice constant,
we can write
\begin{eqnarray}
{\bf R}_{ij\mu\nu}(x,y,z)
&=&
\left[\hat{I}+\hat{\varepsilon}(x,y)\right] \cdot
{\bf R}_{ij\mu\nu}^0,
\end{eqnarray}
where ${\bf R}_{ij\mu\mu}^0$ is the relative interatomic distance
in the absence of strain.
Therefore the hopping integrals will be modified as
\begin{eqnarray}
V_{i,j,\mu,\nu}
\left[R_{ij\mu\nu}(x,y)\right]
&\approx&
V_{i,j,\mu,\nu}
\left(R_{ij\mu\nu}^0\right)\nonumber\\
&\times&\left[
1
-
\beta_{i,j,\mu,\nu}\frac{\delta R_{ij\mu\nu}(x,y)}{ R_{ij\mu\nu}^0}
\right],
\end{eqnarray}
where $\delta R_{ij\mu\nu}(x,y)=R_{ij\mu\nu}(x,y)-R_{ij\mu\nu}^0$.
The deformation coupling parameter
$\beta_{i,j,\mu,\nu}=-d\log V_{i,j,\mu,\nu}(R)/ d\log R$
characterizes the electron-phonon interaction strength
corresponding to each hopping bond, which can be evaluated microscopically from first principle techniques. 

It is illustrative to consider the modification of the band structure of a MoS$_2$ ribbon with a non-uniform profile of strain, originated from delamination from the substrate \cite{A84}. For that aim, it is useful to consider a one-dimensional profile along the $y$-axis of the strain tensor. Neglecting, as a first approximation, the change in the membrane thickness with the in-plane applied strain, one can write a strain tensor
\begin{eqnarray}
\hat{\varepsilon}(x,y,z)
&=&
\hat{\varepsilon}(y)
=
\varepsilon_{\rm max}f(y)
\left(
\begin{array}{ccc}
-{\tilde\nu} & 0 & 0 \\
0 & 1 & 0 \\
0 & 0 & 0
\end{array}
\right),
\label{Eq:StrainNonUniform}
\end{eqnarray}
where $L$ is the characteristic width of the ribbon,
$\varepsilon_{\rm max}$ the maximum strain attained at the
center of the ribbon $y=L/2$, and $\tilde\nu$ is the Poisson's ratio
of the membrane and $f(y)$ is the function that accounts for the distribution of strain along the sample. For a half-sinusoidal distribution we have $f(y)=\sin\left(\frac{\pi y}{L}\right)$, but other distributions of the strain along the sample, like Lorentzian distribution, Gaussian distribution, etc., can be considered with this model \cite{CR12}. The results are shown in Fig. \ref{Fig:InhomogTMDs}.  Panel (a) shows the band structure obtained for different levels of strain. One observes that the direct band gap is reduced by application of strain, in agreement with experimental observations, as discussed in Sec. \ref{Sec:BandGap}. The local change in the gap due to non-uniform strain is shown in a remarkable way by the {\it bending} of the local density of states (LDOS) along the wrinkle, as shown in Fig. \ref{Fig:InhomogTMDs}(b). The plot shows how the gap is reduced at the center of the ribbon, for which the strain is maximum, as it is seen by the cuts shown in Fig. \ref{Fig:InhomogTMDs}(c).

\subsubsection{Gauge fields and topological aspects}

Local modulation of the strain can be used to mimic gauge fields in the samples \cite{GKG10}. This opens the way for studying more fundamental phenomena in 2D crystals, for example, realizing giant pseudo-magnetic fields and associated zero-field quantum Hall effects as observed in graphene \cite{LC10}.

Cazalilla {\it et al.} have suggested in Ref. \onlinecite{COG14} that time-reversal invariant topological phases and quantum spin Hall effect can be induced in single-layer TMDs as MoS$_2$ (at low carrier densities) due to the lack of inversion symmetry and SOC. Indeed, semiconducting TMDs under shear strain are expected to develop spin-polarized Landau levels in different valleys. For MoS$_2$, gaps between consecutive Landau levels have been estimated to scale as $\hbar\omega_c/ k_B \sim 2.7 B_0[\rm T]$ K, in terms of the cyclotron frequency $\omega_c=eB_0/m^*$, where $m^*$ is the effective mass, and $B_0[\rm T]$ is the strength of the pseudomagnetic field in units of Tesla. Taking into account that, for graphene, pseudomagnetic fields of the order of $\sim 100$ T have been already observed experimentally \cite{LC10}, and considering the intrinsic limitations imposed by the maximum tensile strength and the value of the shear modulus of these materials, gaps between Landau level of up to $\sim 20$ K are within experimental reach. A different route to tune the electronic topological properties is the application of an electric field, as discussed in \cite{QL14}. For a comprehensive review of effects of strain in graphene and other 2D crystals, from a more fundamental point of view, we refer the reader to B. Amorim {\it et al.} \cite{AV15}.    
\vspace{.1cm}

\section{Conclusions and outlook}\label{Sec:Conclusions}

We reviewed the effect of strain on the new families of semiconducting 2D crystals. The main parts of the review are devoted to the techniques to fabricate and induce strain in these materials, and how strain engineering can be used to modify in a controlled manner their optical and electronics properties. Especial attention has been paid to transition metal dichalcogenides (MoS$_2$, WS$_2$, etc.), since the reported experiments are only limited to this family of materials so far. Several proposal to use band gap engineering for optoelectronics and photovoltaics applications have been discussed. In particular, the use of localized strain can make it possible to fabricate  solar energy funnels which can be used for photovoltaic cells that absorb a broad band of the solar spectrum. 

Apart from the modification of the electronic band structure, straintronics can exploit the piezoelectric properties of some families of 2D crystals, converting mechanical to electrical energy.  Piezoelectricity is a well-studied effect in which stretching or compressing a material causes it to generate an electrical voltage, or the reverse, in which an applied voltage leads to contraction or expansion of the material. MoS$_2$, a material which is not piezoelectric in its bulk configuration, has been shown to become piezoelectric when it is thinned down to a single atomic layer (or to an odd number of them), and flexing it in the proper direction, as it was shown in Ref. \cite{WW14}. Interestingly, the authors monitored the conversion of mechanical to electrical energy, and observed voltage and current outputs,  finding that the output voltage changed sign when the direction of applied strain is reversed, as expected due to lack of inversion symmetry in single layer TMDs. 

Since 2D crystals can be stretched much farther than conventional materials, these observations can open the door to development of new applications for these materials and its unique properties. 
It is also worth mentioning that, although in this review we focused on the tunability of electronic and optical properties,
strain-induced effects can be employed also to tune alternative degree of freedoms, as magnetism \cite{HP14,Manchanda15}
or lattice dynamics \cite{Jiang14}.

\acknowledgments

R.R. acknowledges financial support from the Juan de la Cierva Program, and the Spanish Ministry of Economy (MINECO) through Grant No. FIS2011-23713. A. C-G. acknowledges financial support by the European Union (FP7) through the FP7-Marie Curie Project PIEF-GA-2011-300802 (STRENGTHNANO) and by the Fundacion BBVA through the fellowship ÒI Convocatoria de Ayudas Fundacion BBVA a Investigadores, Innovadores y Creadores CulturalesÓ (ÒSemiconductores ultradelgados: hacia la optoelectronica flexibleÓ). 
E.C. acknowledges support from the European project FP7-PEOPLE-2013-CIG "LSIE 2D", Italian National MIUR Prin project 20105ZZTSE, and the Italian MIUR program "Progetto Premiale 2012" Project ABNANOTECH. F.G. acknowledges the European Union Seventh Framework Programme under grant agreement n604391 Graphene Flagship, and the European Research Council Advanced Grant program (contract 290846).


\begin{thebibliography}{250}
\expandafter\ifx\csname natexlab\endcsname\relax\def\natexlab#1{#1}\fi
\expandafter\ifx\csname bibnamefont\endcsname\relax
  \def\bibnamefont#1{#1}\fi
\expandafter\ifx\csname bibfnamefont\endcsname\relax
  \def\bibfnamefont#1{#1}\fi
\expandafter\ifx\csname citenamefont\endcsname\relax
  \def\citenamefont#1{#1}\fi
\expandafter\ifx\csname url\endcsname\relax
  \def\url#1{\texttt{#1}}\fi
\expandafter\ifx\csname urlprefix\endcsname\relax\def\urlprefix{URL }\fi
\providecommand{\bibinfo}[2]{#2}
\providecommand{\eprint}[2][]{\url{#2}}




\bibitem{A1}Novoselov, K.S., Geim, A.K., Morozov, S.V., Jiang, D.,  Zhang, Y., Dubonos, S.V., Grigorieva, I.V. and Firsov, A.A., {\it Science} {\bf 306}, 666 (2004).
\bibitem{A2}Novoselov, K.S., Jiang, D., Schedin, F., Booth, T.J., Khotkevich, V.V., Morozov, S.V. and Geim, A.K. {\it Proc. Natl. Acad. Sci.} {\bf 102}, 10451 (2005).
\bibitem{A3}Wang, Q.H., Kalantar-Zadeh, K., Kis, A., Coleman, J.N. and Strano, M.S. {\it Nat. Nanotechnol.} {\bf 7}, 699 (2012).
\bibitem{A4}Butler, S.Z. {\it et al.} {\it ACS Nano} {\bf 7}, 2898 (2013).
\bibitem{A5}Xu, M., Liang, T., Shi, M. and Chen, H.  {\it Chem. Rev.} {\bf 113}, 3766 (2013).
\bibitem{A6}Coleman, J.N. {\it et al.} {\it Science} {\bf 331}, 568 (2011).
\bibitem{A7}Staley, N.E., Wu, J., Eklund, P., Liu, Y., Li, L. and Xu, Z. {\it Phys. Rev. B} {\bf 80}, 184505 (2009).
\bibitem{A8}Zhang, H., Liu, Ch.-X., Qi, X.-L., Dai, X., Fang, Z. and Zhang, S.-Ch. {\it Nat. Phys.} {\bf 5}, 438 (2009).
\bibitem{A9}Castellanos-Gomez, A., Agra\"{i}t, N. and Rubio-Bollinger, G. {\it Appl. Phys. Lett.} {\bf 96}, 213116 (2010).
\bibitem{A10}Dean, C.R., Young, A.F., Meric, I., Lee, C., Wang, L., Sorgenfrei, S., Watanabe, K., Taniguchi, T., Kim, P., Shepard, K.L. and Hone, J. {\it Nat. Nanotechnol.} {\bf 5}, 722 (2010).
\bibitem{A11}Castellanos-Gomez, A., Wojtaszek, M., Tombros, N., Agra{\"\i}t, N., van Wees, B.J. and Rubio-Bollinger, G. {\it Small} {\bf 7}, 2491 (2011).
\bibitem{A12}Castellanos-Gomez, A., Navarro-Moratalla, E., Mokry, G., Quereda, J., Pinilla-Cienfuegos, E., Agra{\"\i}t, N., van der Zant, H.S.J., Coronado, E., Steele, G.A. and Rubio-Bollinger, G. {\it Nano Res.} {\bf 5}, 550 (2012).
\bibitem{A13}El-Bana, M. S. {\it et al.} {\it Supercond. Sci. Technol.} {\bf 26}, 125020 (2013).
\bibitem{A14}Radisavljevic, B., Radenovic, A., Brivio, J., Giacometti, V. and Kis, A. {\it Nat. Nanotechnol.} {\bf 6}, 147 (2011).
\bibitem{A15}Han, M., …zyilmaz, B., Zhang, Y. and Kim, P. {\it Phys. Rev. Lett.} {\bf 98}, 206805 (2007).
\bibitem{A16}Moreno-Moreno, M., Castellanos-Gomez, A., Rubio-Bollinger, G., Gomez-Herrero, J. and Agra\"{i}t, N.  {\it Small} {\bf 5}, 924 (2009).
\bibitem{A17}Li, X., Wang, X., Zhang, L., Lee, S. and Dai, H. {\it Science} {\bf 319}, 1229 (2008).
\bibitem{A18}	Oostinga, J. B., Heersche, H. B., Liu, X., Morpurgo, A. F. and Vandersypen, L. M. K. {\it Nat. Mater.} {\bf 7}, 151 (2008).
\bibitem{A19}	Castro, E.V., Novoselov, K.S., Morozov, S.V., Peres, N.M.R., Dos Santos, J.M.B.L., Nilsson, J., Guinea, F., Geim, A.K. and Castro Neto, A.H.  {\it Phys. Rev. Lett.} {\bf 99}, 216802 (2007).
\bibitem{A20}	Balog, R. {\it et al.} {\it Nat. Mater.} {\bf 9}, 315 (2010).
\bibitem{A21}	Elias, D. C. {\it et al.} {\it Science} {\bf 323}, 610 (2009).
\bibitem{A22}	Liu, Y., Xu, F., Zhang, Z., Penev, E. S. and Yakobson, B. I. {\it Nano Lett.} {\bf 14} 6782 (2014).
\bibitem{A23}	Castellanos-Gomez, A., Wojtaszek, M., Arramel, Tombros, N. and van Wees, B. J. {\it Small} {\bf 8}, 1607 (2012).


\bibitem{Pedersen08} Pedersen, T.G. {\it et al.} {\it Phys. Rev. Lett.} {\bf 100}, 136804 (2008).
\bibitem{Kim10} Kim, M. {\it et al.}, {\it Nano Lett.} {\bf 10}, 1125 (2010).
\bibitem{Yuan13} Yuan, S. {\it et al.} {\it Phys. Rev. B} {\bf 87}, 085430 (2013).



\bibitem{A24}	Splendiani, A. et al.  {\it Nano Lett.} {\bf 10}, 1271 (2010).
\bibitem{A25}	Mak, K., Lee, C., Hone, J., Shan, J. and Heinz, T.  {\it Phys. Rev. Lett.} {\bf 105}, 136805 (2010).



\bibitem{A26}	Koppens, F. H. L. {\it et al.}  {\it Nat. Nanotechnol.} {\bf 9}, 780 (2014).
\bibitem{A27}	Xia, F., Wang, H., Xiao, D., Dubey, M. and Ramasubramaniam, A. {\it Nat. Photonics} {\bf 8}, 899 (2014)
\bibitem{A28}	Fiori, G. {\it et al.}  {\it Nat. Nanotechnol.} {\bf 9}, 768 (2014).
\bibitem{A29}	Liu, L., Kumar, S. B., Ouyang, Y. and Guo, J. {\it IEEE Trans. Electron Devices} {\bf 58}, 3042 (2011).
\bibitem{A30}	Ghatak, S., Pal, A. N. and Ghosh, A. {\it ACS Nano} {\bf 5}, 7707 (2011).
\bibitem{A31}	Li, H. {\it et al.} {\it Small} {\bf 8}, 63 (2012).

\bibitem{A33}	Radisavljevic, B., Whitwick, M. B. and Kis, A. {\it ACS Nano} {\bf 5}, 9934 (2011).
\bibitem{A34}	Das, S., Chen, H.-Y., Penumatcha, A. V. and Appenzeller, J. {\it Nano Lett.} {\bf 13}, 100 (2013).
\bibitem{A35}	Lu, P., Wu, X., Guo, W. and Zeng, X. C., {\it Phys. Chem. Chem. Phys.} {\bf 14}, 13035 (2012).
\bibitem{A36}	Yin, Z. {\it et al.}, {\it ACS Nano} {\bf 6}, 74 (2011).
\bibitem{A37}	Wang, H. {\it et al.}, {\it Nano Lett.} {\bf 12}, 4674 (2012).


\bibitem{CS15} Castellanos-Gomez, A., Singh, V., van der Zant, H.S.J. and Steele, G.A., {\it Annalen der Physik} {\bf 527}, 27 (2015)


\bibitem{A38}	Bertolazzi, S., Brivio, J. and Kis, A., {\it ACS Nano} {\bf 5}, 9703 (2011).
\bibitem{A39}	Castellanos-Gomez, A. {\it et al.}, {\it Adv. Mater.} {\bf 24}, 772 (2012).
\bibitem{A40}	Griffith, A. A., {\it Philos. Trans. R. Soc. London} {\bf 221}, 63 (1921).
\bibitem{A41}	Cooper, R. {\it et al.}, {\it Phys. Rev. B} {\bf 87}, 035423 (2013).
\bibitem{A42}	Castellanos-Gomez, A., van der Zant, H. S. J. and Steele, G. A., {\it Nano Res.} {\bf 7}, 572 (2015).
\bibitem{A43}	Tang, D.-M. {\it et al.}, {\it Nat. Commun.} {\bf 5}, 3631 (2014).

\bibitem{LG15} L\'opez-Pol\'{i}n, G., G\'omez-Navarro, C.,  Parente, V., Guinea, F.,  Katsnelson, M. I.,  P\'erez-Murano, F., and G\'omez-Herrero, J., {\it Nat. Physics} {\bf 11} 26 (2015)

\bibitem{RK10} Rold\'an, R., Fasolino, A.,  Zakharchenko, K. V. and Katsnelson, M. I. {\it Phys. Rev. B} {\bf 83}, 174104 (2010)
\bibitem{LopezPolin15} L\'opez-Pol\'{i}n, G. {\it et al.}, arXiv:1504.05521 (2015)

\bibitem{isaacs}
E.B. Isaacs and C.A. Marianetti,
{\it Phys. Rev. B} {\bf 89}, 184111 (2014).


\bibitem{A44}	Scalise, E., Houssa, M., Pourtois, G., AfanasÕev, V. and Stesmans, A.  {\it Nano Res.} {\bf 5}, 43 (2011).
\bibitem{A45}	Feng, J., Qian, X., Huang, C.-W. and Li, J. {\it Nat. Photonics} {\bf 6}, 866 (2012).
\bibitem{A46}	Ghorbani-Asl, M., Borini, S., Kuc, A. and Heine, T., {\it Phys. Rev. B} {\bf 87}, 235434 (2013).
\bibitem{A47}	Mungu\'{i}a, J., Bremond, G., Bluet, J. M., Hartmann, J. M. and Mermoux, M., {\it Appl. Phys. Lett.} {\bf 93}, 102101 (2008).
\bibitem{A48}	Bonaccorso, F. {\it et al.}, {\it Mater. Today} {\bf 15}, 564 (2012).
\bibitem{A49}	Huang, X., Zeng, Z. and Zhang, H., {\it Chem. Soc. Rev.} {\bf 42}, 1934 (2013).



\bibitem{Blake07} Blake, P. {\it et al.}, {\it Appl. Phys. Lett.} {\bf 91}, 063124 (2007).



\bibitem{A50}	Li, H. {\it et al.}, {\it Small} {\bf 8}, 682 (2012)
\bibitem{A51}	Li, H. {\it et al.}, {\it ACS Nano} {\bf 7}, 10344 (2013).
\bibitem{A52}	Benameur, M. M. {\it et al.}, {\bf 22}, 125706 (2011).
\bibitem{A53}	Castellanos-Gomez, A. {\it et al.}, {\it Nano Res.} {\bf 6}, 191 (2013).
\bibitem{A54}	Dols-Perez, A., Sisquella, X., Fumagalli, L. and Gomila, G., {\it Nanoscale Res. Lett.} {\bf 8}, 305 (2013).
\bibitem{A55}	Zhan, Y., Liu, Z., Najmaei, S., Ajayan, P. M. and Lou, J., {\it Small}, {\bf 8}, 966, (2012).
\bibitem{A56}	Najmaei, S. {\it et al.}, {\it Nat. Mater.} {\bf 12}, 754 (2013).
\bibitem{A57}	Van der Zande, A. M. {\it et al.}, {\it Nat. Mater.} {\bf 12}, 554 (2013).
\bibitem{A58}	Lee, Y.-H. {\it et al.}, {\it Nano Lett.} {\bf 13}, 1852 (2013).
\bibitem{A59}	Lee, Y.-H. {\it et al.}, {\it Adv. Mater.} {\bf 24}, 2320 (2012).
\bibitem{A60}	Elias, A. L. {\it et al.}, {\it ACS Nano} {\bf 7}, 5235 (2013).
\bibitem{A61}	Lin, M. {\it et al.}, {\it J. Am. Chem. Soc.} {\bf 135}, 13274 (2013).
\bibitem{A62}	Shi, Y. {\it et al.}, {\it Nano Lett.} {\bf 12}, 2784 (2012).
\bibitem{A63}	Lee, G.-H. {\it et al.}, {\it APL Mater.} {\bf 2}, 092511 (2014).
\bibitem{A64}	Nicolosi, V., Chhowalla, M., Kanatzidis, M. G., Strano, M. S. and Coleman, J. N., {\it Science} {\bf 340}, 1226419 (2013).
\bibitem{A65}	Coleman, J. N. {\it et al.}, {\it Science} {\bf 331}, 568 (2011).
\bibitem{A66}	Smith, R. J. {\it et al.}, {\it Adv. Mater.} {\bf 23}, 3944 (2011).
\bibitem{A67}	Hanlon, D. {\it et al.} arXiv:1501.01881 (2015).
\bibitem{A68}	Yasaei, P. {\it et al.}, {\it Adv. Mater.} 27, 1887 (2015).
\bibitem{A69}	Eda, G. {\it et al.}, {\it Nano Lett.} {\bf 11}, 5111 (2011).
\bibitem{A70}	Zheng, J. {\it et al.}, {\it Nat. Commun.} {\bf 5}, 2995 (2014).
\bibitem{A71}	Ugural, A. C. {\it Mechanics of Materials.} (Wiley, 2008). 
\bibitem{A72}	Mohiuddin, T. {\it et al.}, {\it Phys. Rev. B} {\bf 79}, 205433 (2009).
\bibitem{A73}	Huang, M., Yan, H., Heinz, T. F. and Hone, J., {\it Nano Lett.} {\bf 10}, 4074 (2010).
\bibitem{A74}	Yu, T. {\it et al.}, {\it J. Phys. Chem. C} {\bf 112}, 12602 (2008).
\bibitem{A75}	Ni, Z. H. {\it et al.}, {\it ACS Nano} {\bf 3}, 483 (2009).
\bibitem{A76}	Ni, Z. H. {\it et al.}  {\it ACS Nano} {\bf 2}, 2301 (2008).
\bibitem{A77}	Conley, H. J. {\it et al.}, {\it Nano Lett.} {\bf 13}, 3626 (2013).
\bibitem{A78}	He, K., Poole, C., Mak, K. F. and Shan, J., {\it Nano Lett.} {\bf 13}, 2931 (2013).
\bibitem{A79}	Zhu, C. R. {\it et al.}, {\it Phys. Rev. B} {\bf 88}, 121301 (2013).
\bibitem{A80}	Wang, Y., Cong, C., Yang, W., Shang, J. and Peimyoo, N., {\it Nano Res.} (2015). doi:10.1007/s12274-015-0762-6
\bibitem{A81}	Hui, Y. Y. {\it et al.}, {\it ACS Nano} {\bf 7}, 7126 (2013).
\bibitem{A82}	Plechinger, G. {\it et al.}, {\it Phys. Status Solidi - Rapid Res. Lett.} {\bf 3}, 126, (2012).
\bibitem{A83}	Plechinger, G. {\it et al.} {\it 2D Mater.} {\bf 2}, 015006 (2015).


\bibitem{Zang13} Zang, J. {\it et al.} {\it Nat. Materials} {\bf 12}, 321 (2013)
\bibitem{Wang11} Wang, Y. {\it et al.} {\it ACS Nano} {\bf 5}, 3645 (2011)


\bibitem{A84}	Castellanos-Gomez, A. {\it et al.}, {\it Nano Lett.} {\bf 13}, 5361 (2013).
\bibitem{A85}	Yang, S. {\it et al.}, {\it Nano Lett.} {\bf 15}, 1660 (2015).
\bibitem{A86}	Vella, D., Bico, J., Boudaoud, A., Roman, B. and Reis, P. M. {\it Proc. Natl. Acad. Sci.} {\bf 106}, 10901 (2009).
\bibitem{A87}	Rice, C. {\it et al.}, {\it Phys. Rev. B} {\bf 87}, 081307 (2013).
\bibitem{A88}	Desai, S. B. {\it et al.}, {\it Nano Lett.} {\bf 14}, 4592 (2014).
\bibitem{A89}	Yan, R. {\it et al.}, arXiv:1211.4136 (2012). 


\bibitem{KZH11} Kuc, A., Zibouche, N., and Heine, T., {\it Phys. Rev. B} {\bf 83}, 245213 (2011) 



\bibitem[{\citenamefont{Liu et~al.}(2013)\citenamefont{Liu, Shan, Yao, Yao, and
  Xiao}}]{LX13}
\bibinfo{author}{\bibfnamefont{G.-B.} \bibnamefont{Liu}},
  \bibinfo{author}{\bibfnamefont{W.-Y.} \bibnamefont{Shan}},
  \bibinfo{author}{\bibfnamefont{Y.}~\bibnamefont{Yao}},
  \bibinfo{author}{\bibfnamefont{W.}~\bibnamefont{Yao}}, \bibnamefont{and}
  \bibinfo{author}{\bibfnamefont{D.}~\bibnamefont{Xiao}},
  \bibinfo{journal}{{\it Phys. Rev. B}} \textbf{\bibinfo{volume}{88}},
  \bibinfo{pages}{085433} (\bibinfo{year}{2013}).

\bibitem[{\citenamefont{Ko\ifmmode~\acute{s}\else \'{s}\fi{}mider
  et~al.}(2013)\citenamefont{Ko\ifmmode~\acute{s}\else \'{s}\fi{}mider,
  Gonz\'alez, and Fern\'andez-Rossier}}]{KGF13}
\bibinfo{author}{\bibfnamefont{K.}~\bibnamefont{Ko\ifmmode~\acute{s}\else
  \'{s}\fi{}mider}}, \bibinfo{author}{\bibfnamefont{J.~W.}
  \bibnamefont{Gonz\'alez}}, \bibnamefont{and}
  \bibinfo{author}{\bibfnamefont{J.}~\bibnamefont{Fern\'andez-Rossier}},
  \bibinfo{journal}{{\it Phys. Rev. B}} \textbf{\bibinfo{volume}{88}},
  \bibinfo{pages}{245436} (\bibinfo{year}{2013}).


\bibitem[{\citenamefont{Bromley et~al.}(1972)\citenamefont{Bromley, Murray, and
  Yoffe}}]{BMY72}
\bibinfo{author}{\bibfnamefont{R.~A.} \bibnamefont{Bromley}},
  \bibinfo{author}{\bibfnamefont{R.~B.} \bibnamefont{Murray}},
  \bibnamefont{and} \bibinfo{author}{\bibfnamefont{A.~D.} \bibnamefont{Yoffe}},
  \bibinfo{journal}{{\it J. Phys. C: Solid State Phys.}}
  \textbf{\bibinfo{volume}{5}}, \bibinfo{pages}{759} (\bibinfo{year}{1972}).
  
  
\bibitem{SBJ87}  Schutte, W.,  Boer, J. and Jellinek, F. {\it J. Solid State Chem.} {\bf 70}, 207 (1987)

\bibitem[{\citenamefont{Kumar and Ahluwalia}(2012)}]{KA12}
\bibinfo{author}{\bibfnamefont{A.}~\bibnamefont{Kumar}} \bibnamefont{and}
  \bibinfo{author}{\bibfnamefont{P.}~\bibnamefont{Ahluwalia}},
  \bibinfo{journal}{{\it The European Physical Journal B}}
  \textbf{\bibinfo{volume}{85}}, \bibinfo{pages}{1} (\bibinfo{year}{2012}).

  
  

\bibitem[{\citenamefont{Rold{\'a}n et~al.}(2014)\citenamefont{Rold{\'a}n,
  Silva-Guill{\'e}n, L{\'o}pez-Sancho, Guinea, Cappelluti, and
  Ordej{\'o}n}}]{RO14}
\bibinfo{author}{\bibfnamefont{R.}~\bibnamefont{Rold{\'a}n}},
  \bibinfo{author}{\bibfnamefont{J.~A.} \bibnamefont{Silva-Guill{\'e}n}},
  \bibinfo{author}{\bibfnamefont{M.~P.} \bibnamefont{L{\'o}pez-Sancho}},
  \bibinfo{author}{\bibfnamefont{F.}~\bibnamefont{Guinea}},
  \bibinfo{author}{\bibfnamefont{E.}~\bibnamefont{Cappelluti}},
  \bibnamefont{and}
  \bibinfo{author}{\bibfnamefont{P.}~\bibnamefont{Ordej{\'o}n}},
  \bibinfo{journal}{{\it Annalen der Physik}} \textbf{\bibinfo{volume}{526}},
  \bibinfo{pages}{347} (\bibinfo{year}{2014}).
  
\bibitem[{\citenamefont{Rold\'an et~al.}(2013)\citenamefont{Rold\'an,
  Cappelluti, and Guinea}}]{RCG13}
\bibinfo{author}{\bibfnamefont{R.}~\bibnamefont{Rold\'an}},
  \bibinfo{author}{\bibfnamefont{E.}~\bibnamefont{Cappelluti}},
  \bibnamefont{and} \bibinfo{author}{\bibfnamefont{F.}~\bibnamefont{Guinea}},
  \bibinfo{journal}{{\it Phys. Rev. B}} \textbf{\bibinfo{volume}{88}},
  \bibinfo{pages}{054515} (\bibinfo{year}{2013}).


\bibitem[{\citenamefont{Rudenko and Katsnelson}(2014)}]{RK14}
\bibinfo{author}{\bibfnamefont{A.~N.} \bibnamefont{Rudenko}} \bibnamefont{and}
  \bibinfo{author}{\bibfnamefont{M.~I.} \bibnamefont{Katsnelson}},
  \bibinfo{journal}{{\it Phys. Rev. B}} \textbf{\bibinfo{volume}{89}},
  \bibinfo{pages}{201408} (\bibinfo{year}{2014})

\bibitem[{\citenamefont{Liu et~al.}(2011)\citenamefont{Liu, Feng, and
  Yao}}]{LFY11}
\bibinfo{author}{\bibfnamefont{C.-C.} \bibnamefont{Liu}},
  \bibinfo{author}{\bibfnamefont{W.}~\bibnamefont{Feng}}, \bibnamefont{and}
  \bibinfo{author}{\bibfnamefont{Y.}~\bibnamefont{Yao}},
  \bibinfo{journal}{{\it Phys. Rev. Lett.}} \textbf{\bibinfo{volume}{107}},
  \bibinfo{pages}{076802} (\bibinfo{year}{2011}).
  
  \bibitem{Tongay14} Tongay, S. {\it et al.} {\it Nat. Comm.} {\bf 5}, 3252




\bibitem[{\citenamefont{Cappelluti et~al.}(2013)\citenamefont{Cappelluti,
  Rold\'an, Silva-Guill\'en, Ordej\'on, and Guinea}}]{CG13}
\bibinfo{author}{\bibfnamefont{E.}~\bibnamefont{Cappelluti}},
  \bibinfo{author}{\bibfnamefont{R.}~\bibnamefont{Rold\'an}},
  \bibinfo{author}{\bibfnamefont{J.~A.} \bibnamefont{Silva-Guill\'en}},
  \bibinfo{author}{\bibfnamefont{P.}~\bibnamefont{Ordej\'on}},
  \bibnamefont{and} \bibinfo{author}{\bibfnamefont{F.}~\bibnamefont{Guinea}},
  \bibinfo{journal}{{\it Phys. Rev. B}} \textbf{\bibinfo{volume}{88}},
  \bibinfo{pages}{075409} (\bibinfo{year}{2013}).

\bibitem[{\citenamefont{Rold\'an et~al.}(2014)\citenamefont{Rold\'an,
  L\'opez-Sancho, Guinea, Cappelluti, Silva-Guill\'en, and
  Ordej\'on}}]{RO14-2D}
\bibinfo{author}{\bibfnamefont{R.}~\bibnamefont{Rold\'an}},
  \bibinfo{author}{\bibfnamefont{M.~P.} \bibnamefont{L\'opez-Sancho}},
  \bibinfo{author}{\bibfnamefont{F.}~\bibnamefont{Guinea}},
  \bibinfo{author}{\bibfnamefont{E.}~\bibnamefont{Cappelluti}},
  \bibinfo{author}{\bibfnamefont{J.~A.} \bibnamefont{Silva-Guill\'en}},
  \bibnamefont{and}
  \bibinfo{author}{\bibfnamefont{P.}~\bibnamefont{Ordej\'on}},
  \bibinfo{journal}{{\it 2D Materials}} \textbf{\bibinfo{volume}{1}},
  \bibinfo{pages}{034003} (\bibinfo{year}{2014})

\bibitem[{\citenamefont{{Livneh} and {Spanier}}(2014)}]{LS14b}
\bibinfo{author}{\bibfnamefont{T.}~\bibnamefont{{Livneh}}} \bibnamefont{and}
  \bibinfo{author}{\bibfnamefont{J.~E.} \bibnamefont{{Spanier}}},
  \bibinfo{journal}{ArXiv e-prints}  (\bibinfo{year}{2014}),
  \eprint{1408.6748}.

\bibitem[{\citenamefont{Tran et~al.}(2014)\citenamefont{Tran, Soklaski, Liang,
  and Yang}}]{TY14}
\bibinfo{author}{\bibfnamefont{V.}~\bibnamefont{Tran}},
  \bibinfo{author}{\bibfnamefont{R.}~\bibnamefont{Soklaski}},
  \bibinfo{author}{\bibfnamefont{Y.}~\bibnamefont{Liang}}, \bibnamefont{and}
  \bibinfo{author}{\bibfnamefont{L.}~\bibnamefont{Yang}},
  \bibinfo{journal}{{\it Phys. Rev. B}} \textbf{\bibinfo{volume}{89}},
  \bibinfo{pages}{235319} (\bibinfo{year}{2014}).

\bibitem[{\citenamefont{Morita}(1986)}]{M86}
\bibinfo{author}{\bibfnamefont{A.}~\bibnamefont{Morita}},
  \bibinfo{journal}{{\it Applied Physics A}} \textbf{\bibinfo{volume}{39}},
  \bibinfo{pages}{227} (\bibinfo{year}{1986}).

\bibitem[{\citenamefont{Seifert and Hern‡ndez}(2000)}]{SH00}
\bibinfo{author}{\bibfnamefont{G.}~\bibnamefont{Seifert}} \bibnamefont{and}
  \bibinfo{author}{\bibfnamefont{E.}~\bibnamefont{Hern‡ndez}},
  \bibinfo{journal}{{\it Chemical Physics Letters}} \textbf{\bibinfo{volume}{318}},
  \bibinfo{pages}{355 } (\bibinfo{year}{2000})

\bibitem[{\citenamefont{Castellanos-Gomez
  et~al.}(2014)\citenamefont{Castellanos-Gomez, Vicarelli, Prada, Island,
  Narasimha-Acharya, Blanter, Groenendijk, Buscema, Steele, Alvarez
  et~al.}}]{CZ14}
\bibinfo{author}{\bibfnamefont{A.}~\bibnamefont{Castellanos-Gomez}},
  \bibinfo{author}{\bibfnamefont{L.}~\bibnamefont{Vicarelli}},
  \bibinfo{author}{\bibfnamefont{E.}~\bibnamefont{Prada}},
  \bibinfo{author}{\bibfnamefont{J.~O.} \bibnamefont{Island}},
  \bibinfo{author}{\bibfnamefont{K.~L.} \bibnamefont{Narasimha-Acharya}},
  \bibinfo{author}{\bibfnamefont{S.~I.} \bibnamefont{Blanter}},
  \bibinfo{author}{\bibfnamefont{D.~J.} \bibnamefont{Groenendijk}},
  \bibinfo{author}{\bibfnamefont{M.}~\bibnamefont{Buscema}},
  \bibinfo{author}{\bibfnamefont{G.~A.} \bibnamefont{Steele}},
  \bibinfo{author}{\bibfnamefont{J.~V.} \bibnamefont{Alvarez}},
  \bibnamefont{et~al.}, \bibinfo{journal}{{\it 2D Materials}}
  \textbf{\bibinfo{volume}{1}}, \bibinfo{pages}{025001} (\bibinfo{year}{2014})
  
 \bibitem{XWJ14} Xia, F., Wang, H. and Jia, Y., {\it Nature Comm.} {\bf 5} 4458 (2014)
  
\bibitem{LG14} Low, T., Rold\'an, R., Wang, H., Xia, F.,  Avouris, P.,  Moreno, L.M.  and Guinea, F. {\it Phys. Rev. Lett.}, {\bf 113} 106802 (2014) 
  
  
  
  
\bibitem[{\citenamefont{Tao et~al.}(2015)\citenamefont{Tao, Cinquanta, Chiappe,
  Grazianetti, Fanciulli, Dubey, Molle, and Akinwande}}]{TA15}
\bibinfo{author}{\bibfnamefont{L.}~\bibnamefont{Tao}},
  \bibinfo{author}{\bibfnamefont{E.}~\bibnamefont{Cinquanta}},
  \bibinfo{author}{\bibfnamefont{D.}~\bibnamefont{Chiappe}},
  \bibinfo{author}{\bibfnamefont{C.}~\bibnamefont{Grazianetti}},
  \bibinfo{author}{\bibfnamefont{M.}~\bibnamefont{Fanciulli}},
  \bibinfo{author}{\bibfnamefont{M.}~\bibnamefont{Dubey}},
  \bibinfo{author}{\bibfnamefont{A.}~\bibnamefont{Molle}}, \bibnamefont{and}
  \bibinfo{author}{\bibfnamefont{D.}~\bibnamefont{Akinwande}},
  \bibinfo{journal}{{\it Nat. Nanotech.}} \textbf{\bibinfo{volume}{10}},
  \bibinfo{pages}{227} (\bibinfo{year}{2015}).


\bibitem[{\citenamefont{Scalise et~al.}(2014)\citenamefont{Scalise, Houssa,
  Pourtois, Afanas?ev, and Stesmans}}]{SS12}
\bibinfo{author}{\bibfnamefont{E.}~\bibnamefont{Scalise}},
  \bibinfo{author}{\bibfnamefont{M.}~\bibnamefont{Houssa}},
  \bibinfo{author}{\bibfnamefont{G.}~\bibnamefont{Pourtois}},
  \bibinfo{author}{\bibfnamefont{V.}~\bibnamefont{Afanas?ev}},
  \bibnamefont{and} \bibinfo{author}{\bibfnamefont{A.}~\bibnamefont{Stesmans}},
  \bibinfo{journal}{{\it Physica E: Low-dimensional Systems and Nanostructures}}
  \textbf{\bibinfo{volume}{56}}, \bibinfo{pages}{416 } (\bibinfo{year}{2014}).

\bibitem[{\citenamefont{Scalise et~al.}(2012)\citenamefont{Scalise, Houssa,
  Pourtois, AfanasÕev, and Stesmans}}]{SS12b}
\bibinfo{author}{\bibfnamefont{E.}~\bibnamefont{Scalise}},
  \bibinfo{author}{\bibfnamefont{M.}~\bibnamefont{Houssa}},
  \bibinfo{author}{\bibfnamefont{G.}~\bibnamefont{Pourtois}},
  \bibinfo{author}{\bibfnamefont{V.}~\bibnamefont{AfanasÕev}},
  \bibnamefont{and} \bibinfo{author}{\bibfnamefont{A.}~\bibnamefont{Stesmans}},
  \bibinfo{journal}{{\it Nano Research}} \textbf{\bibinfo{volume}{5}},
  \bibinfo{pages}{43} (\bibinfo{year}{2012}).

\bibitem[{\citenamefont{Yun et~al.}(2012)\citenamefont{Yun, Han, Hong, Kim, and
  Lee}}]{YL12}
\bibinfo{author}{\bibfnamefont{W.~S.} \bibnamefont{Yun}},
  \bibinfo{author}{\bibfnamefont{S.~W.} \bibnamefont{Han}},
  \bibinfo{author}{\bibfnamefont{S.~C.} \bibnamefont{Hong}},
  \bibinfo{author}{\bibfnamefont{I.~G.} \bibnamefont{Kim}}, \bibnamefont{and}
  \bibinfo{author}{\bibfnamefont{J.~D.} \bibnamefont{Lee}},
  \bibinfo{journal}{{\it Phys. Rev. B}} \textbf{\bibinfo{volume}{85}},
  \bibinfo{pages}{033305} (\bibinfo{year}{2012}).

\bibitem[{\citenamefont{Johari and Shenoy}(2012)}]{JS12}
\bibinfo{author}{\bibfnamefont{P.}~\bibnamefont{Johari}} \bibnamefont{and}
  \bibinfo{author}{\bibfnamefont{V.~B.} \bibnamefont{Shenoy}},
  \bibinfo{journal}{{\it ACS Nano}} \textbf{\bibinfo{volume}{6}},
  \bibinfo{pages}{5449} (\bibinfo{year}{2012}).

\bibitem[{\citenamefont{Pan and Zhang}(2012)}]{PZ12}
\bibinfo{author}{\bibfnamefont{H.}~\bibnamefont{Pan}} \bibnamefont{and}
  \bibinfo{author}{\bibfnamefont{Y.-W.} \bibnamefont{Zhang}},
  \bibinfo{journal}{{\it The Journal of Physical Chemistry C}}
  \textbf{\bibinfo{volume}{116}}, \bibinfo{pages}{11752}
  (\bibinfo{year}{2012}).

\bibitem{zhang2013}
Q. Zhang, Y. Cheng, L.-Y. Gan, and U. Schwingenschl\"ogl,
Phys. Rev. B {\bf 88}, 245447 (2013).

\bibitem{shi13}
H. Shi, H. Pan, Y.-W. Zhang, and B.I. Yakobson,
Phys. Rev. B {\bf 87}, 155304 (2013).


\bibitem{guzman}
D.M. Guzman and A. Strachan,
J. Appl. Phys. {\bf 115}, 243701 (2014).

\bibitem{jiang}
J.-W. Jiang,
Nanoscale {\bf 6}, 8326 (2014).

\bibitem{Horzum2013}
Horzum, S., Sahin, H., Cahangirov, S., Cudazzo, P., Rubio, A., Serin, T. and Peeters, F. M., 
{\it Phys. Rev. B} {\bf 87}, 125415 (2013)


\bibitem{bhatta}
S. Bhattacharyya, T. Pandey, and A.K. Singh,
Nanotechnology {\bf 25}, 465701 (2014).

\bibitem{hosseini}
M. Hosseini, M. Elahi, M. Pourfath, and D. Esseni,
arXiv:1503.01301 (2015).

\bibitem{rhim}
S.H. Rhim and A.J. Freeman,
arXiv:1504.04936 (2015).

\bibitem{wangkutana}
L. Wang, A. Kutana, and B.I. Yakobson,
Ann. Phys. {\bf 526}, L7 (2014).

\bibitem[{\citenamefont{Peelaers and Van~de Walle}(2012)}]{PV12}
\bibinfo{author}{\bibfnamefont{H.}~\bibnamefont{Peelaers}} \bibnamefont{and}
  \bibinfo{author}{\bibfnamefont{C.~G.} \bibnamefont{Van~de Walle}},
  \bibinfo{journal}{{\it Phys. Rev. B}} \textbf{\bibinfo{volume}{86}},
  \bibinfo{pages}{241401} (\bibinfo{year}{2012}).



\bibitem[{\citenamefont{\ifmmode \mbox{\c{C}}\else \c{C}\fi{}ak\ifmmode \imath
  \else~\i \fi{}r et~al.}(2014)\citenamefont{\ifmmode \mbox{\c{C}}\else
  \c{C}\fi{}ak\ifmmode \imath \else~\i \fi{}r, Sahin, and Peeters}}]{CSP14}
\bibinfo{author}{\bibfnamefont{D.}~\bibnamefont{\ifmmode \mbox{\c{C}}\else
  \c{C}\fi{}ak\ifmmode \imath \else~\i \fi{}r}},
  \bibinfo{author}{\bibfnamefont{H.}~\bibnamefont{Sahin}}, \bibnamefont{and}
  \bibinfo{author}{\bibfnamefont{F.~m. c.~M.} \bibnamefont{Peeters}},
  \bibinfo{journal}{{\it Phys. Rev. B}} \textbf{\bibinfo{volume}{90}},
  \bibinfo{pages}{205421} (\bibinfo{year}{2014}).

\bibitem[{\citenamefont{Liu et~al.}(2014)\citenamefont{Liu, Neal, Zhu, Luo, Xu,
  Tom‡nek, and Ye}}]{LY14}
\bibinfo{author}{\bibfnamefont{H.}~\bibnamefont{Liu}},
  \bibinfo{author}{\bibfnamefont{A.~T.} \bibnamefont{Neal}},
  \bibinfo{author}{\bibfnamefont{Z.}~\bibnamefont{Zhu}},
  \bibinfo{author}{\bibfnamefont{Z.}~\bibnamefont{Luo}},
  \bibinfo{author}{\bibfnamefont{X.}~\bibnamefont{Xu}},
  \bibinfo{author}{\bibfnamefont{D.}~\bibnamefont{Tom‡nek}}, \bibnamefont{and}
  \bibinfo{author}{\bibfnamefont{P.~D.} \bibnamefont{Ye}},
  \bibinfo{journal}{{\it ACS Nano}} \textbf{\bibinfo{volume}{8}},
  \bibinfo{pages}{4033} (\bibinfo{year}{2014}).

\bibitem[{\citenamefont{Rodin et~al.}(2014)\citenamefont{Rodin, Carvalho, and
  Castro~Neto}}]{RCC14}
\bibinfo{author}{\bibfnamefont{A.~S.} \bibnamefont{Rodin}},
  \bibinfo{author}{\bibfnamefont{A.}~\bibnamefont{Carvalho}}, \bibnamefont{and}
  \bibinfo{author}{\bibfnamefont{A.~H.} \bibnamefont{Castro~Neto}},
  \bibinfo{journal}{{\it Phys. Rev. Lett.}} \textbf{\bibinfo{volume}{112}},
  \bibinfo{pages}{176801} (\bibinfo{year}{2014}).

\bibitem[{\citenamefont{Lv et~al.}(2014)\citenamefont{Lv, Lu, Shao, and
  Sun}}]{LS14}
\bibinfo{author}{\bibfnamefont{H.~Y.} \bibnamefont{Lv}},
  \bibinfo{author}{\bibfnamefont{W.~J.} \bibnamefont{Lu}},
  \bibinfo{author}{\bibfnamefont{D.~F.} \bibnamefont{Shao}}, \bibnamefont{and}
  \bibinfo{author}{\bibfnamefont{Y.~P.} \bibnamefont{Sun}},
  \bibinfo{journal}{{\it Phys. Rev. B}} \textbf{\bibinfo{volume}{90}},
  \bibinfo{pages}{085433} (\bibinfo{year}{2014}).

\bibitem[{\citenamefont{Appalakondaiah
  et~al.}(2012)\citenamefont{Appalakondaiah, Vaitheeswaran, Leb\`egue,
  Christensen, and Svane}}]{AS12}
\bibinfo{author}{\bibfnamefont{S.}~\bibnamefont{Appalakondaiah}},
  \bibinfo{author}{\bibfnamefont{G.}~\bibnamefont{Vaitheeswaran}},
  \bibinfo{author}{\bibfnamefont{S.}~\bibnamefont{Leb\`egue}},
  \bibinfo{author}{\bibfnamefont{N.~E.} \bibnamefont{Christensen}},
  \bibnamefont{and} \bibinfo{author}{\bibfnamefont{A.}~\bibnamefont{Svane}},
  \bibinfo{journal}{{\it Phys. Rev. B}} \textbf{\bibinfo{volume}{86}},
  \bibinfo{pages}{035105} (\bibinfo{year}{2012}).

\bibitem[{\citenamefont{{Liao} et~al.}(2014)\citenamefont{{Liao}, {Zhou},
  {Qiu}, {Dresselhaus}, and {Chen}}}]{LC14}
\bibinfo{author}{\bibfnamefont{B.}~\bibnamefont{{Liao}}},
  \bibinfo{author}{\bibfnamefont{J.}~\bibnamefont{{Zhou}}},
  \bibinfo{author}{\bibfnamefont{B.}~\bibnamefont{{Qiu}}},
  \bibinfo{author}{\bibfnamefont{M.~S.} \bibnamefont{{Dresselhaus}}},
  \bibnamefont{and} \bibinfo{author}{\bibfnamefont{G.}~\bibnamefont{{Chen}}},
  \bibinfo{journal}{ArXiv e-prints}  (\bibinfo{year}{2014}),
  \eprint{1410.4242}.

\bibitem[{\citenamefont{Guan et~al.}(2014)\citenamefont{Guan, Zhu, and
  Tom\'anek}}]{GZT14}
\bibinfo{author}{\bibfnamefont{J.}~\bibnamefont{Guan}},
  \bibinfo{author}{\bibfnamefont{Z.}~\bibnamefont{Zhu}}, \bibnamefont{and}
  \bibinfo{author}{\bibfnamefont{D.}~\bibnamefont{Tom\'anek}},
  \bibinfo{journal}{{\it Phys. Rev. Lett.}} \textbf{\bibinfo{volume}{113}},
  \bibinfo{pages}{046804} (\bibinfo{year}{2014}).

\bibitem[{\citenamefont{Fei and Yang}(2014)}]{FL14}
\bibinfo{author}{\bibfnamefont{R.}~\bibnamefont{Fei}} \bibnamefont{and}
  \bibinfo{author}{\bibfnamefont{L.}~\bibnamefont{Yang}},
  \bibinfo{journal}{{\it Nano Letters}} \textbf{\bibinfo{volume}{14}},
  \bibinfo{pages}{2884} (\bibinfo{year}{2014}).

\bibitem[{\citenamefont{Wei and Peng}(2014)}]{WP14}
\bibinfo{author}{\bibfnamefont{Q.}~\bibnamefont{Wei}} \bibnamefont{and}
  \bibinfo{author}{\bibfnamefont{X.}~\bibnamefont{Peng}},
  \bibinfo{journal}{{\it Applied Physics Letters}} \textbf{\bibinfo{volume}{104}},
  \bibinfo{eid}{251915} (\bibinfo{year}{2014})

\bibitem[{\citenamefont{Peng et~al.}(2014)\citenamefont{Peng, Wei, and
  Copple}}]{PWC14}
\bibinfo{author}{\bibfnamefont{X.}~\bibnamefont{Peng}},
  \bibinfo{author}{\bibfnamefont{Q.}~\bibnamefont{Wei}}, \bibnamefont{and}
  \bibinfo{author}{\bibfnamefont{A.}~\bibnamefont{Copple}},
  \bibinfo{journal}{{\it Phys. Rev. B}} \textbf{\bibinfo{volume}{90}},
  \bibinfo{pages}{085402} (\bibinfo{year}{2014}).

\bibitem[{\citenamefont{Elahi et~al.}(2015)\citenamefont{Elahi, Khaliji,
  Tabatabaei, Pourfath, and Asgari}}]{EA15}
\bibinfo{author}{\bibfnamefont{M.}~\bibnamefont{Elahi}},
  \bibinfo{author}{\bibfnamefont{K.}~\bibnamefont{Khaliji}},
  \bibinfo{author}{\bibfnamefont{S.~M.} \bibnamefont{Tabatabaei}},
  \bibinfo{author}{\bibfnamefont{M.}~\bibnamefont{Pourfath}}, \bibnamefont{and}
  \bibinfo{author}{\bibfnamefont{R.}~\bibnamefont{Asgari}},
  \bibinfo{journal}{{\it Phys. Rev. B}} \textbf{\bibinfo{volume}{91}},
  \bibinfo{pages}{115412} (\bibinfo{year}{2015}).

\bibitem[{\citenamefont{Ge et~al.}(2015)\citenamefont{Ge, Wan, Yang, and
  Yao}}]{GY15}
\bibinfo{author}{\bibfnamefont{Y.}~\bibnamefont{Ge}},
  \bibinfo{author}{\bibfnamefont{W.}~\bibnamefont{Wan}},
  \bibinfo{author}{\bibfnamefont{F.}~\bibnamefont{Yang}}, \bibnamefont{and}
  \bibinfo{author}{\bibfnamefont{Y.}~\bibnamefont{Yao}}, \bibinfo{journal}{{\it New
  Journal of Physics}} \textbf{\bibinfo{volume}{17}}, \bibinfo{pages}{035008}
  (\bibinfo{year}{2015})
  
  
\bibitem[{\citenamefont{Cai et~al.}(2013)\citenamefont{Cai, Chuu, Wei, and
  Chou}}]{CC13}
\bibinfo{author}{\bibfnamefont{Y.}~\bibnamefont{Cai}},
  \bibinfo{author}{\bibfnamefont{C.-P.} \bibnamefont{Chuu}},
  \bibinfo{author}{\bibfnamefont{C.~M.} \bibnamefont{Wei}}, \bibnamefont{and}
  \bibinfo{author}{\bibfnamefont{M.~Y.} \bibnamefont{Chou}},
  \bibinfo{journal}{{\it Phys. Rev. B}} \textbf{\bibinfo{volume}{88}},
  \bibinfo{pages}{245408} (\bibinfo{year}{2013}).

\bibitem[{\citenamefont{Kaloni and Schwingenschl{\"o}gl}(2013)}]{KS13}
\bibinfo{author}{\bibfnamefont{T.}~\bibnamefont{Kaloni}} \bibnamefont{and}
  \bibinfo{author}{\bibfnamefont{U.}~\bibnamefont{Schwingenschl{\"o}gl}},
  \bibinfo{journal}{{\it Chemical Physics Letters}} \textbf{\bibinfo{volume}{583}},
  \bibinfo{pages}{137} (\bibinfo{year}{2013}).

\bibitem[{\citenamefont{Wang and Ding}(2013)}]{WD13}
\bibinfo{author}{\bibfnamefont{Y.}~\bibnamefont{Wang}} \bibnamefont{and}
  \bibinfo{author}{\bibfnamefont{Y.}~\bibnamefont{Ding}},
  \bibinfo{journal}{{\it Solid State Communications}} \textbf{\bibinfo{volume}{155}},
  \bibinfo{pages}{6 } (\bibinfo{year}{2013})
  
  

\bibitem[{\citenamefont{Feng et~al.}(2012)\citenamefont{Feng, Qian, Huang, and
  Li}}]{FL12}
\bibinfo{author}{\bibfnamefont{J.}~\bibnamefont{Feng}},
  \bibinfo{author}{\bibfnamefont{X.}~\bibnamefont{Qian}},
  \bibinfo{author}{\bibfnamefont{C.-W.} \bibnamefont{Huang}}, \bibnamefont{and}
  \bibinfo{author}{\bibfnamefont{J.}~\bibnamefont{Li}},
  \bibinfo{journal}{{\it Nature Photonics}} \textbf{\bibinfo{volume}{6}},
  \bibinfo{pages}{866} (\bibinfo{year}{2012}).
  
  
  
  
  
  
  
  
\bibitem{WQL14}  Wu, M., Qian, X. and Li, J.,{\it Nano Lett.}, {\bf 14}, 5350 (2014)
  
  \bibitem{Mehboudi15} Mehboudi, M. {\it et al.}, {\it PNAS}  doi:10.1073/pnas.1500633112 (2015) 
  
  \bibitem{Neek2013} Neek-Amal, M. and Beheshtian, J. and Sadeghi, A. and Michel, K. H. and Peeters, F. M., {\it The Journal of Physical Chemistry C}, {\bf 117}, 13261 (2013)
  
  

\bibitem[{\citenamefont{Slater and Koster}(1954)}]{SK54}
\bibinfo{author}{\bibfnamefont{J.~C.} \bibnamefont{Slater}} \bibnamefont{and}
  \bibinfo{author}{\bibfnamefont{G.~F.} \bibnamefont{Koster}},
  \bibinfo{journal}{{\it Phys. Rev.}} \textbf{\bibinfo{volume}{94}},
  \bibinfo{pages}{1498} (\bibinfo{year}{1954}).

\bibitem[{\citenamefont{Castro-Neto et~al.}(2009)\citenamefont{Castro-Neto,
  Guinea, Peres, Novoselov, and Geim}}]{CG09}
\bibinfo{author}{\bibfnamefont{A.~H.} \bibnamefont{Castro-Neto}},
  \bibinfo{author}{\bibfnamefont{F.}~\bibnamefont{Guinea}},
  \bibinfo{author}{\bibfnamefont{N.~M.~R.} \bibnamefont{Peres}},
  \bibinfo{author}{\bibfnamefont{K.}~\bibnamefont{Novoselov}},
  \bibnamefont{and} \bibinfo{author}{\bibfnamefont{A.~K.} \bibnamefont{Geim}},
  \bibinfo{journal}{{\it Rev. Mod. Phys.}} \textbf{\bibinfo{volume}{81}},
  \bibinfo{pages}{109} (\bibinfo{year}{2009}).

\bibitem[{\citenamefont{Vozmediano et~al.}(2010)\citenamefont{Vozmediano,
  Katsnelson, and Guinea}}]{VKG10}
\bibinfo{author}{\bibfnamefont{M.~A.} \bibnamefont{Vozmediano}},
  \bibinfo{author}{\bibfnamefont{M.}~\bibnamefont{Katsnelson}},
  \bibnamefont{and} \bibinfo{author}{\bibfnamefont{F.}~\bibnamefont{Guinea}},
  \bibinfo{journal}{{\it Physics Reports}} \textbf{\bibinfo{volume}{496}},
  \bibinfo{pages}{109} (\bibinfo{year}{2010}).

\bibitem[{\citenamefont{Castellanos-G{\'o}mez
  et~al.}(2012)\citenamefont{Castellanos-G{\'o}mez, Poot, Steele, van~der Zant,
  Agra{\"\i}t, and Rubio-Bollinger}}]{CR12}
\bibinfo{author}{\bibfnamefont{A.}~\bibnamefont{Castellanos-G{\'o}mez}},
  \bibinfo{author}{\bibfnamefont{M.}~\bibnamefont{Poot}},
  \bibinfo{author}{\bibfnamefont{G.~A.} \bibnamefont{Steele}},
  \bibinfo{author}{\bibfnamefont{H.~S.~J.} \bibnamefont{van~der Zant}},
  \bibinfo{author}{\bibfnamefont{N.}~\bibnamefont{Agra{\"\i}t}},
  \bibnamefont{and}
  \bibinfo{author}{\bibfnamefont{G.}~\bibnamefont{Rubio-Bollinger}},
  \bibinfo{journal}{{\it Adv. Mater.}} \textbf{\bibinfo{volume}{24}},
  \bibinfo{pages}{772} (\bibinfo{year}{2012}).

\bibitem[{\citenamefont{Guinea et~al.}(2010)\citenamefont{Guinea, Katsnelson,
  and Geim}}]{GKG10}
\bibinfo{author}{\bibfnamefont{F.}~\bibnamefont{Guinea}},
  \bibinfo{author}{\bibfnamefont{M.}~\bibnamefont{Katsnelson}},
  \bibnamefont{and} \bibinfo{author}{\bibfnamefont{A.}~\bibnamefont{Geim}},
  \bibinfo{journal}{{\it Nature Physics}} \textbf{\bibinfo{volume}{6}},
  \bibinfo{pages}{30} (\bibinfo{year}{2010}).

\bibitem[{\citenamefont{Levy et~al.}(2010)\citenamefont{Levy, Burke, Meaker,
  Panlasigui, Zettl, Guinea, Neto, and Crommie}}]{LC10}
\bibinfo{author}{\bibfnamefont{N.}~\bibnamefont{Levy}},
  \bibinfo{author}{\bibfnamefont{S.}~\bibnamefont{Burke}},
  \bibinfo{author}{\bibfnamefont{K.}~\bibnamefont{Meaker}},
  \bibinfo{author}{\bibfnamefont{M.}~\bibnamefont{Panlasigui}},
  \bibinfo{author}{\bibfnamefont{A.}~\bibnamefont{Zettl}},
  \bibinfo{author}{\bibfnamefont{F.}~\bibnamefont{Guinea}},
  \bibinfo{author}{\bibfnamefont{A.~C.} \bibnamefont{Neto}}, \bibnamefont{and}
  \bibinfo{author}{\bibfnamefont{M.}~\bibnamefont{Crommie}},
  \bibinfo{journal}{{\it Science}} \textbf{\bibinfo{volume}{329}},
  \bibinfo{pages}{544} (\bibinfo{year}{2010}).

\bibitem[{\citenamefont{Cazalilla et~al.}(2014)\citenamefont{Cazalilla, Ochoa,
  and Guinea}}]{COG14}
\bibinfo{author}{\bibfnamefont{M.~A.} \bibnamefont{Cazalilla}},
  \bibinfo{author}{\bibfnamefont{H.}~\bibnamefont{Ochoa}}, \bibnamefont{and}
  \bibinfo{author}{\bibfnamefont{F.}~\bibnamefont{Guinea}},
  \bibinfo{journal}{{\it Phys. Rev. Lett.}} \textbf{\bibinfo{volume}{113}},
  \bibinfo{pages}{077201} (\bibinfo{year}{2014}).

\bibitem[{\citenamefont{Qian et~al.}(2014)\citenamefont{Qian, Junwei, Fu, and
  Li}}]{QL14}
\bibinfo{author}{\bibfnamefont{X.}~\bibnamefont{Qian}},
  \bibinfo{author}{\bibnamefont{Junwei}},
  \bibinfo{author}{\bibfnamefont{L.}~\bibnamefont{Fu}}, \bibnamefont{and}
  \bibinfo{author}{\bibfnamefont{J.}~\bibnamefont{Li}},
  \bibinfo{journal}{{\it Science}} \textbf{\bibinfo{volume}{346}},
  \bibinfo{pages}{1344} (\bibinfo{year}{2014}).
  
  \bibitem{AV15} B. Amorim, A. Cortijo, F. de Juan, A. G. Grushin, F. Guinea, A. Guti\'errez-Rubio, H. Ochoa, V. Parente, R. Rold\'an, P. San-Jos\'e, J. Schiefele, M. Sturla, and M. A. H. Vozmediano, arXiv:1503.00747 (2015).
  
  
  
 

\bibitem[{\citenamefont{Wang and Wu}(2014)}]{WW14}
\bibinfo{author}{\bibfnamefont{L.}~\bibnamefont{Wang}} \bibnamefont{and}
  \bibinfo{author}{\bibfnamefont{M.}~\bibnamefont{Wu}},
  \bibinfo{journal}{{\it Physics Letters A}} \textbf{\bibinfo{volume}{378}},
  \bibinfo{pages}{1336 } (\bibinfo{year}{2014}).
 

\bibitem{HP14}
H. Pan, 
{\it Sci. Rep.} {\bf 4}, 7524 (2014).


\bibitem{Manchanda15}
P. Manchanda, V. Sharma, H. Yu, D. J. Sellmyer, and R Skomski,
arXiv:1504.04602 (2015)

\bibitem{Jiang14}
J.-W. Jiang, 
{\it Nanoscale} {\bf 6}, 8326 (2014).

  
  

\end{thebibliography}

\end{document}